\documentclass[aps,prx,superscriptaddress]{revtex4-2}
\usepackage{graphicx, xcolor}
\usepackage{amsmath}
\usepackage{braket}

\newcommand{\figref}[1]{\figurename~\ref{#1}}
\newcommand{\secref}[1]{Sec.~\ref{#1}}
\renewcommand{\eqref}[1]{Eq.~(\ref{#1})}

\newcommand{\ip}{I_\mathrm{p}}
\newcommand{\up}{U_\mathrm{p}}
\newcommand{\pb}{\mathbf{p}}

\newcommand{\kb}{\mathbf{k}}

\newcommand{\Ab}{\mathbf{A}}

\newcommand{\xhat}{\hat{\bf x}}
\newcommand{\yhat}{\hat{\bf y}}

\begin{document}

	\title{Manipulating Twisted Electrons in Strong-Field Ionization}
	
	\author{A. S. Maxwell}
	\email{andrew.maxwell@ucl.ac.uk}	
	\affiliation{Department of Physics \& Astronomy, University College London, Gower Street, London WC1E 6BT, United Kingdom}
	\affiliation{ICFO---Institut de Ciencies Fotoniques, The Barcelona Institute of Science and Technology, 08860 Castelldefels (Barcelona), Spain}
	
	\author{G. S. J. Armstrong}
	\affiliation{Centre for Theoretical Atomic, Molecular and Optical Physics, School of Mathematics and Physics, Queen’s University Belfast, University Road, Belfast BT7 1NN, Northern Ireland, United Kingdom}
	
	\author{M. F. Ciappina}
	\affiliation{ICFO---Institut de Ciencies Fotoniques, The Barcelona Institute of Science and Technology, 08860 Castelldefels (Barcelona), Spain}
	\affiliation{Physics Program, Guangdong Technion – Israel Institute of Technology, Shantou, Guangdong 515063, China}
	\affiliation{Technion – Israel Institute of Technology, Haifa, 32000, Israel}
	
	\author{E. Pisanty}
	\affiliation{ICFO---Institut de Ciencies Fotoniques, The Barcelona Institute of Science and Technology, 08860 Castelldefels (Barcelona), Spain}
	
	\author{Y. Kang}
	\affiliation{Department of Physics \& Astronomy, University College London, Gower Street, London WC1E 6BT, United Kingdom}
	
	\author{A. C. Brown}
	\affiliation{Centre for Theoretical Atomic, Molecular and Optical Physics, School of Mathematics and Physics, Queen’s University Belfast, University Road, Belfast BT7 1NN, Northern Ireland, United Kingdom}

	\author{M. Lewenstein}
	\affiliation{ICFO---Institut de Ciencies Fotoniques, The Barcelona Institute of Science and Technology, 08860 Castelldefels (Barcelona), Spain}
	\affiliation{ICREA, Pg. Lluís Companys 23, 08010}
	
	\author{C. Figueira de Morisson Faria}
	\affiliation{Department of Physics \& Astronomy, University College London, Gower Street, London WC1E 6BT, United Kingdom}

	\date{\today}
	\begin{abstract}
		We investigate the discrete orbital angular momentum (OAM) of photoelectrons freed in strong-field ionization. We use these `twisted' electrons to  provide an alternative interpretation on existing experimental work of vortex interferences caused by strong field ionization mediated by two counter-rotating circularly polarized pulses separated by a delay.
		Using the strong field approximation, we derive an interference condition for the vortices. In computations for a neon target we find very good agreement of the vortex condition with photoelectron momentum distributions computed with the strong field approximation, as well as with the time-dependent methods Qprop and R-Matrix.
		For each of these approaches we examine the OAM of the photoelectrons, finding a small number of vortex states localized in separate energy regions.		
		We demonstrate that the vortices arise from the interference of pairs of twisted electron states. 		
		The OAM of each twisted electron state can be directly related to the number of arms of the spiral in that region.
		We gain further understanding by recreating the vortices with pairs of twisted electrons and use this to determine a semiclassical relation for the OAM.
		A discussion is included on measuring the OAM in strong field ionization directly or by employing specific laser pulse schemes as well as utilizing the OAM in time-resolved imaging of photo-induced dynamics.
	\end{abstract}

	\maketitle

	\section{Introduction}
	
	Phase vortices in light and matter are the product of the orbital angular momentum (OAM) carried by free particles resulting in a rotating wavefront. The OAM of a free particle is a quantum number that can be measured and manipulated, with many of the same properties as spin, yet it has only been observed in experiments fairly recently in photons (3 decades ago) \cite{Allen1992} and even later for electrons (1 decade ago) \cite{Bliokh2007} (see \cite{Bliokh2017,Lloyd2017} for reviews). It has been found that these \emph{twisted} photons or electrons have huge potential in imaging, and recent experiments with electron vortex beams have revealed their role in chiral energy-loss spectroscopy and magnetic dichroism \cite{Bliokh2017}.

	The study of twisted light and electrons in strong field processes is a very new but rapidly developing topic. In high-harmonic generation (HHG) it has been shown that a twisted driving field can be used to produce twisted UV light \cite{Zurch2012}. Later this was made time-varying \cite{Rego2019} and further extended to the generation of UV light with torus knot topology \cite{Pisanty2019}. The effect of twisted electrons has been investigated in HHG \cite{Gemsheim2019}. However, the effect of OAM on photoelectrons in strong-field ionization has not been studied in detail.
	
	Recently, the OAM content of photoelectrons was computed using the strong field approximation (SFA). This work focused on reaching exceedingly high OAM values for quasi-relativistic field intensities \cite{Velez2018} and terahertz fields \cite{Velez2020}. 
	Twisted electrons have also been studied in the context of rescattering with a well defined OAM \cite{Tolstikhin2019}.
	However, currently there is very little work focused on the generation, measurement and manipulation of photoelectrons carrying OAM for strong field ionization.	
	That said, there is a still a lot to learn from related strong field work. Through conservation of momentum, photoelectrons ionized by a linearly polarized laser field will adopt the same OAM as the magnetic quantum number of the bound state of origin \cite{Tolstikhin2019}. The linear field does not alter the magnetic quantum number or the photoelectron's OAM, and there is a direct mapping between the two.
	Furthermore, it has been demonstrated that circularly polarized laser fields preferentially tunnel ionize $p$ electrons, whose magnetic quantum number $m$ dictates that they are `counter-rotating' with respect to the laser field \cite{Barth2011,Barth2013}. Depending on the helicity of the laser field, either $m=1$ or $m=-1$ electrons dominate the signal, while ionization from orbitals with $m=0$ is strongly suppressed. Thus, given the interaction with the magnetic quantum number, tunnel ionization via circularly polarized light may be a route to generate, control, and measure photoelectrons with non-trivial OAM.
	
	A more explicit example of twisted electrons in multiphoton processes is the formation of interference vortices \cite{NgokoDjiokap2015,NgokoDjiokap2016,yuan2016,NgokoDjiokap2018,Pengel2017,Pengel2017a,Li2018, Kerbstadt2019,Kerbstadt2019a, Armstrong2019b}. Computations of the momentum distribution of photoelectrons ionized via two time-delayed, counter-rotating circularly-polarized laser fields reveal a Fermat spiral interference pattern. Initially treated theoretically for single-photon ionization \cite{NgokoDjiokap2015} and later two-photon ionization \cite{NgokoDjiokap2016} of helium, such spirals have since been observed in both single and double ionization of molecules \cite{yuan2016,NgokoDjiokap2018}. Experimental verification came in Refs.\;\cite{Pengel2017, Pengel2017a}, where the spiral interference pattern was demonstrated for three-photon ionization of potassium. Calculations using the multielectron \textit{ab initio} $R$-Matrix with time dependence (RMT) method fully corroborated this effect \cite{Armstrong2019b}, and revealed the three-dimensional characteristics of the interference vortices. Experimental progress in this domain continues to gather pace, and vortex measurements at mid-infrared wavelengths are being carried out \cite{bayer2020}.
	Interference vortices are indicative of photoelectrons carrying OAM, as the interference of two states with differing OAM produces such spirals. The above-mentioned  multiphoton studies have provided explanations in terms of the magnetic quantum number in the bound states and scattering states to which the photoelectron is promoted, but do not go as far as discussing the OAM of the free electron or its measurement and control. In this work we examine the interference vortices in the non-perturbative regime, where a description of promotion to scattering states will not hold. However, analysis of co- and counter-rotating electrons \cite{Barth2011,Barth2013} is valid and provides additional insight.
	
	Twisted electrons and the OAM distributions are the focus of this article, and with these tools we present an alternative understanding of the vortices. 
	Using the SFA \cite{amini2019}, as well as the time-dependent Schr\"odinger equation (TDSE)-based solvers, Qprop~\cite{qprop2,qprop3} and RMT \cite{moore2011,clarke2018,rmtcpc}, we compute both the photoelectron's momentum distribution and its constituent OAM components. The resulting momentum-space vortices may be understood by deriving an interference condition using the SFA, and by recreating the same pattern using the interference between states of different values of OAM. We also identify a semi-classical relationship that demonstrates a link between the OAM and above threshold ionization peaks of the photoelectron.
	Understanding the OAM in photoelectrons, which undergo strong-field ionization via circularly-polarized light, imbues the seemingly plain momentum distributions with previously unseen structure revealed only by such interferometric schemes. This opens the question: is there a generalised way to measure, directly or indirectly, the OAM of photoelectrons in strong-field experiments? 
	Given the high correlation between the photoelectron OAM and the quantum numbers of the initial, bound state, consideration of the OAM could be extremely useful in relating physical observables to the bound state and disentangling interferences in time-resolved measurements. Furthermore, the potential for the outgoing twisted photoelectron states to be chiral could aid the detection and spectrographic measurement of chiral molecular targets \cite{Asenjo-Garcia2014}.
	
	The article is structured as follows. In section \secref{Sec:Methods} we outline the theoretical background for the three methods employed in this article, the SFA (\secref{Sec:Methods:SFA}), the single active electron (SAE) 3D-TDSE solver Qprop  (\secref{Sec:Methods:Qprop}), and the multielectron TDSE solver RMT (\secref{Sec:Methods:RMT}). 
	In \secref{Sec:Methods:OAM} we present the methodology for computing the OAM distributions for each model. Next we focus on the interference vortices themselves. In \secref{Sec:Vortex}, we present the derivation of the vortex interference condition using the SFA (\secref{Sec:VortexCondition}) and compare the results of all methods and that condition (\secref{Sec:MomentumDistribusions}). While in \secref{Sec:OAM} we turn our attention to the orbital angular momentum (OAM), in \secref{Sec:OAMDistrbusions} we compare the photoelectron OAM distribution from all methods and in \secref{Sec:OAMSpiral} we use this to construct the interference vortex entirely from twisted electron states.
	In \secref{Sec:Conclusions} we make our concluding statements and discuss potential direct and indirect measurement schemes. We employ atomic units throughout (denoted a.u.), where the elementary charge, electron mass and the Planck constant are set to one, $e=m=\hbar=1$.

	\section{Background and Methods}
	\label{Sec:Methods}
	
	\subsubsection{Laser field}
	\label{sec:LaserPulse}
	We consider a bicircular field comprised of two time-delayed, 400-nm pulses, described by the vector potential
	\begin{align}
	    \Ab(t) &=\Ab_+(t)+\Ab_{-}(t-\delta),\\
	    \intertext{where $\pm$ denotes the rotation of the circular field. Each field is given by}
	    \Ab_{\pm}(t) = & \ \frac{A_0}{\sqrt{2}} \sin^2 \left(\frac{\omega t}{2N_c}\right)
	    \left[
	         \sin\omega t \;\xhat \pm \cos\omega t \;\yhat
	         \right],
	\end{align}
	where $A_0=2\sqrt{\up}$ is the peak vector potential strength, which is related to the peak electric field amplitude by $A_0=E_0/\omega$. The laser frequency is $\omega = 0.114$ a.u. (corresponding to a 400~nm wavelength), $N_c$ is the total number of laser cycles, and $\delta$ is the time delay between the pulses, which we will typically take to be $\delta=\frac{2\pi N_c}{\omega}$ so that the two field envelopes do not overlap.
	Each pulse reaches an associated laser peak intensity of $2.5\times10^{13}$ W/cm$^2$ and the field retains this profile for time $t\in[0,2\pi N_c/\omega]$, and is zero otherwise.
	
	\subsection{Strong Field Approximation}
	\label{Sec:Methods:SFA}

	Our starting point in the SFA \cite{amini2019} is the transition amplitude for direct ATI from the initial bound state $|\psi_0\rangle$ to a final Volkov state with drift momentum \textbf{p} given by \cite{Becker2002Review, Faria2002, Keldysh1965, *Faisal1973, *Reiss1980}
	\begin{equation}
	M(\pb)=-i\lim\limits_{t\to\infty}  e^{i S(\pb,t)} \int_{-\infty}^{t}dt' d(\pb,t') e^{i S(\pb,t')},
	\label{eq:SFA-transition-general}
	\end{equation}
	where $d(\pb,t')=\braket{ \pb+\Ab(t')|V|\Psi_{0}}$, $\ket{\Psi_{0}}$ is the ground state of the target and the action is given by
	\begin{equation}
		S(\mathbf{p},t)=\ip t +\frac{1}{2}\int d t (\pb+\Ab(t))^2.
		\label{eq:SFA-action-general}
	\end{equation}
    Here, $\ip$ is the ionization potential of our target.
    We employ the saddle point approximation, seeking stationary action for the integration variable $t'$,
    \begin{equation}
    	\frac{\partial S}{\partial t'}=\ip+\frac{1}{2}\left(\pb+\Ab(t_s)\right)^2=0.
    	\label{eq:SFA-Times}
    \end{equation}
    Now the probability distribution can be computed from \eqref{eq:SFA-transition-general} as
    \begin{align}
    	M(\pb)=&
    	\sum_s 
       	c(\pb,t_s,t)	d(\pb,t_s) e^{i S(\pb,t_s)},
    	\intertext{where the prefactor $c(\pb,t_s, t)$, derived from application of the saddle point approximation and also includes the $t'$ independent phase from \eqref{eq:SFA-transition-general},  is given by}
    	c(\pb,t_s,t)&=-i e^{i S(\pb,t)}\sqrt{\frac{2\pi i}{\partial^2 S(\pb,t_s)/\partial t_s^2}}.
    \end{align}

    In order to capture the essential physics of the interference vortices, we make some additional assumptions. The minimal requirement is to have interference between two photoelectron orbits deriving from each pulse. Thus, in the SFA model we take only two ionization events from near the peak of each laser pulse, $\Ab_{+}(t)$ and $\Ab_{-}(t-\delta)$, but otherwise neglect the pulse envelopes to allow for an analytic description. Thus, in the SFA model, the vector potentials are defined by
    \begin{equation}
        \Ab_{\pm}(t)=\frac{A_0}{\sqrt{2}}\left(\sin(\omega t)\xhat 
        \pm \cos(\omega t)\yhat\right),
    \end{equation}
    where the $\pm$ denotes the direction of rotation of the field.
    In this way we are able to separate the two contributions of ionization at the times $t_+$ and $t_-+\delta$, given by
    \begin{align}
        \omega t_{\pm} = \mp\phi
        -\arcsin\left( 
            \frac{(2 \ip+p^2+2 \up)\csc(\theta)}{2p\sqrt{2\up}}
        \right).
        \label{eq:times}
    \end{align}
    In \eqref{eq:times} we have written the solutions in spherical momentum coordinates $p$, $\theta$ and $\phi$ as they are the most natural for this system.
    The transition amplitude can then be written as a sum of the two separate solutions
    \begin{align}
        M(\pb)&=
         c_{+}(\pb,t)\;
         d_{+}(\pb)\exp(i S_{+}(\pb)) 
         +c_{-}(\pb,t)\;
            d_{-}(\pb)\exp(i S_{-}(\pb)).
    \label{eq:SFA-Vortex-Amplitude}
    \end{align}
    The actions $S_{+}(\pb)$, $\;S_{-}(\pb)$  and prefactors $c_{+}(\pb,t)$, $c_{-}(\pb,t)$, $d_{+}(\pb)$ and $d_{-}(\pb)$ have the times $t_{+}$ or $t_{-}+\delta$ substituted into them such that
    \begin{align}
        S_{+}(\pb)=\ip t_{+}+\frac{1}{2}\int_{t_{+}}d\tau (\pb+\Ab_{+}(\tau))^2, &&
        S_{-}(\pb)=\ip (t_{-}+\delta)+\frac{1}{2}\int_{t_{-}+\delta}d\tau (\pb+\Ab_{-}(\tau-\delta))^2
    \end{align}
    and
    \begin{align}
        d_{+}(\pb)&=d(\pb,t_{+})&
        d_{-}(\pb)&=d(\pb,t_{-}+\delta).\notag\\
        c_{+}(\pb,t)&=c(\pb,t_{+},t)&
        c_{-}(\pb,t)&=c(\pb,t_{-}+\delta,t).&
	\end{align}
	Following the prescription of \cite{Barth2011} we split the prefactor $d(\pb,t)$ into two parts
	\begin{equation}
	    d(\pb,t)=e^{i m \phi_k(\pb,t)} f_{nlm}(k(\pb,t), \theta_k(\pb,t)),
	    \label{eq:Dipole-Factoring}
	\end{equation}
	where $\kb=\pb+\Ab(t)$ and $k(\pb,t)$, $ \theta_k(\pb,t)$ and $\phi_k(\pb,t)$ are the spherical coordinates of $\kb$.
	The function $f_{nlm}(k,\theta)$ is dependent on the specific bound state, which in our case is the ground state of neon, while $\phi_k$ is the so-called the tunneling angle \cite{Barth2011}, which leads to the enhanced ionization of `counter-rotating' electrons. It will continue to play an important role in the description of the interference vortices.

	\subsection{Qprop}
	\label{Sec:Methods:Qprop}
	The 3D-TDSE calculations were performed using the latest version of Qprop~\cite{qprop3}, where a very accurate method for the calculation of photoelectron spectra (PES), dubbed i-SURFV, is implemented. In short, Qprop is a velocity gauge 3D-TDSE solver that allows studies within both the single active electron (SAE) approximation and many-electron systems via the solution of the time-dependent Kohn–Sham equations. From a computational viewpoint, the velocity gauge formulation of the strong laser-matter problem appears to be more efficient, providing results of accuracy equivalent to the length gauge with much less computational effort. For our SAE model of the Ne atom, we have employed the model potential of Ref.~\cite{tong} and started the time-dependent simulations from the 2$p_0$, 2$p_{-1}$ and 2$p_{+1}$ bound states. These states were computed via imaginary time propagation.
	The pulse parameters are those given in \secref{sec:LaserPulse} and, specifically, the effective potential has the form
	\begin{equation}
	    V(r)=-\frac{Z+f(r)}{r}\quad \text{with}\quad f(r)=a_1 e^{-a_2 r}+a_3 r e^{-a_4 r} + a_5 e^{-a_6 r},
	\end{equation}
	where $Z=1$. For neon the coefficients $a_i$ are $a_1=8.069$, $a_2=2.148$, $a_3=-3.570$, $a_4=1.986$, $a_5=0.931$, $a_6=0.602$~\cite{tong}, which gives the correct ionization potential of $\ip=0.79$~a.u. In this computation we considered angular momenta up to $L=14$ and all magnetic sublevels.

	\subsection{R-Matrix with time dependence}
	\label{Sec:Methods:RMT}
	
	The $R$-matrix with time dependence method (RMT) is an {\em ab initio} approach that solves the time-dependent Schr\"{o}dinger equation for single ionization of multielectron atoms, ions, and molecules in arbitrarily-polarized, strong laser fields \cite{moore2011,clarke2018,rmtcpc}. The method divides position space into two distinct regions. The inner region, which usually extends to around 20 Bohr radii (20 a.u.), confines the target nucleus. In this region, a many-body wave function is treated, accounting for electron exchange and correlation. The outer region, which extends to larger distances, contains a single ejected electron that interacts with both the laser field and the singly-ionized residual target. 
	
	Our description of the neon target is provided in a previous work \cite{burke1975}. In short, we couple a single electron to the Ne$^+$ residual ion, expanding the wave function in $LM_LS\pi$ symmetries up to $L=99$, retaining both the \(1s^2 2s^2 2p^5\) and \(1s^2 2s 2p^6\) Ne$^+$ residual ion states. The photoelectron wave function is calculated on a finite-difference radial grid in the outer region, which extends to 3400 a.u.. 
	The laser pulse configuration is that given in \secref{sec:LaserPulse}.
	The wave function is propagated in time for a total of 35 fs, which gives sufficient time for the ionizing wavepacket to extend to large distances within the outer region. The photoelectron momentum distribution is obtained by performing a Fourier transform of the ejected-electron radial wavefunction.
	
	\subsection{Computing OAM Distributions}
	\label{Sec:Methods:OAM}
	In order to compute the orbital angular momentum distributions we will take two approaches. Firstly, given a transition amplitude such as that in the SFA we can transform it into the `OAM basis' by computing its Fourier series coefficients \cite{Velez2018}
	\begin{equation}
	    M_{l_v}(p_{||},p_{\perp})=
	    \frac{1}{2\pi}
	    \int_{-\pi}^{\pi} d \phi\; e^{-i l_{v} \phi} M(\pb).
	    \label{eq:OAMTransformation}
	\end{equation}
	However, given that this is a transformation into cylindrical co-ordinates, and many TDSE solvers employ a basis of spherical harmonics, some computational power can be saved by converting a distribution expanded in such a basis,  
	\begin{equation}
	    M(\pb) = \sum_{l=0}^{L}\sum_{m=-l}^{l}M_{lm}(p)Y_l^m(\theta,\phi),
	\end{equation}
	so that the transformation of \eqref{eq:OAMTransformation} reduces to
	\begin{equation}
	    M_{l_v}(p_{||},p_{\perp})=
	    \sum^{L}_{l=|l_v|}  M_{l l_{v}}(p)Y^{l_v}_{l}(\theta,0),
	    \label{eq:OAMSphericalTransformation}
	\end{equation}
	where $\tan\theta = p_{\perp}/p_{||}$, and $L$ is the maximum value of the azimuthal quantum number $l$ required for adequate convergence.
	
	\section{Vortex interference}
	\label{Sec:Vortex}
	
    \subsection{Vortex interference condition from the standard SFA implementation}
    \label{Sec:VortexCondition}
    
    In this section we derive a condition for the interference vortices from the SFA description. Such conditions can help to give some insight into the physical dynamics of the process.
    \begin{figure}
		\includegraphics[width=0.8\linewidth]{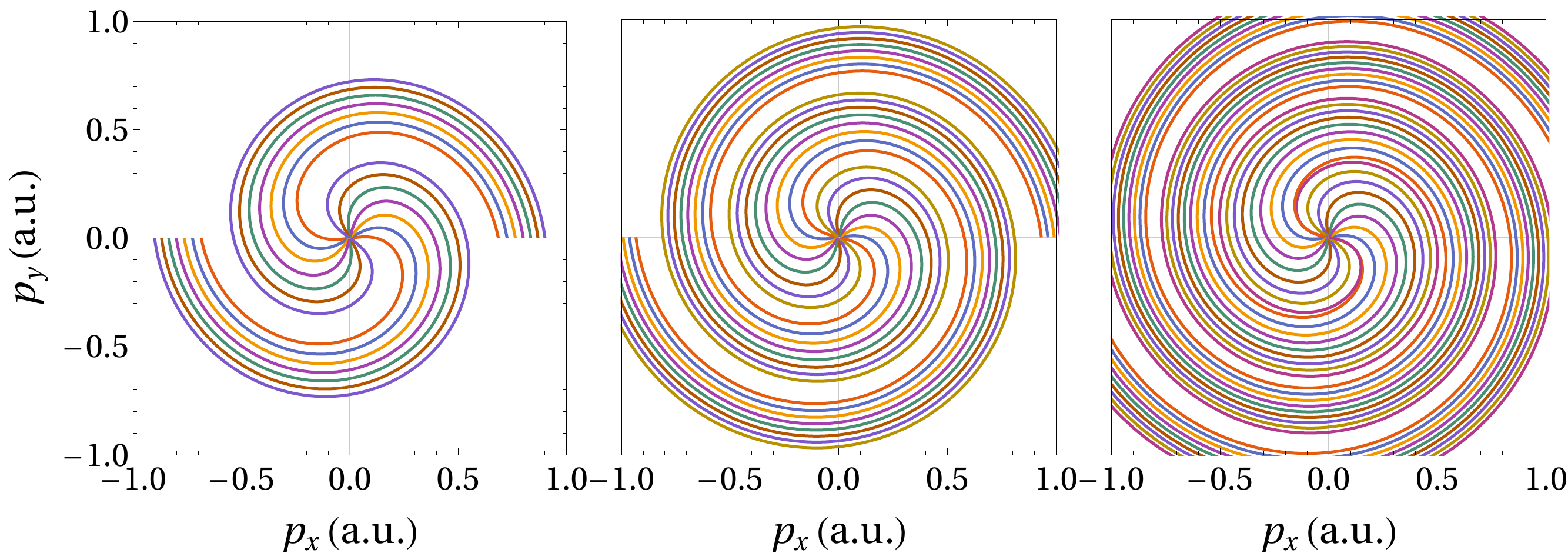}
		\caption{Plotting the interference vortex condition over the polarization $p_x p_y$-plane of the laser field. The three spirals are plotted using 3 different sets of $n$ from \eqref{eq:VortexCondition}; the first 7 valid $n$, the next 8 and then the next 9 are used, from left to right. The minimum valid value of $n$ is 37 for our field, target parameters and $\phi=\pi$.
		Using a delay $\delta=4\times 2\pi/\omega$, an angular frequency of $\omega = 0.114$~a.u., $\up=0.027$~a.u., corresponding to a peak laser intensity of $I_0=2.5\times 10^{13}$~W/cm$^2$. The ionization potential is $\ip=0.79$~a.u. for a neon target.
		} 
		\label{fig:SpiralCond}
	\end{figure}
    The starting point to derive the condition is \eqref{eq:SFA-Vortex-Amplitude}. We can demonstrate that the prefactors $c_{+}(\pb,t)$ and $c_{-}(\pb,t)$ are the same and given by
    \begin{equation}
        c_{\pm}(\pb, t)=\frac{i e^{i S(\pb,t)}}{2}\sin(\theta)\omega
                        \sqrt{8\up p^2-(2\ip+2\up+p^2)\csc(\theta)}.
    \end{equation}
    Thus, $c_{\pm}(\pb, t)$ can be factored out and will not play a role in the interference.
    For the prefactor $d_{\pm}(\pb)$,
    co-ordinates $k$ and $\theta_k$ of $f_{nlm}(k, \theta_k)$ can be shown to be the same when evaluated at either $t_{+}$ or $t_{-}+\delta$. From the saddle point \eqref{eq:SFA-Times} we can determine
    \begin{align}
        k(\pb,t_{+})^2&=k(\pb,t_{-}+\delta)^2=-2\ip \notag\\
        \cos(\theta_k(\pb,t_{+}))&=\cos(\theta_k(\pb,t_{-}+\delta))=-\frac{p}{2\ip}\cos(\theta).
    \end{align}
    Thus, we can deduce
    \begin{align}
        f_{nlm}\left(k(\pb,t_{+}),\theta_k(\pb,t_{+}) \right)= 
        f_{nlm}\left(k(\pb,t_{-}+\delta),\theta_k(\pb,t_{-}+\delta) \right)=:f_{nlm}(k,\theta_k).
    \end{align}
    Hence, $f_{nlm}(k,\theta_k)$ can be factored out and we are left with
    the $e^{i m \phi_k}$ term, which (as presented in \cite{Barth2011}) will act to weight the contribution from each laser pulse depending on the value of $m$. For non-zero $m$ this will lead to a loss of contrast of the interference fringes as electrons `co-rotating' with the field will be preferentially ionized. Now we may write the transition amplitude as
    \begin{equation}
    M(\pb)= C e^{i m \phi_{k+}(\pb)} e^{i S_+(\pb)}
    \left( 1 + e^{i m \Delta \phi_k(\pb)}  e^{i \Delta S(\pb)} \right),
    \end{equation}
    where $C$ contains all the prefactors. The term $e^{i m \Delta \phi_k(\pb)}$ acts to switch off or blur the interference in the way described.
    The interference condition can be derived by maximizing the term $e^{i \Delta S(\pb)}$, where
    \begin{equation}
        \Delta S(\pb) = \frac{
            (2 \ip +p^2 +2\up)(2\phi+\delta \omega).
        }{2\omega}
    \end{equation}
    Setting $\Delta S(\pb) = 2\pi n$, where $n$ is an integer we can find its maxima and thus a condition for the interference vortices
	\begin{equation}
	    \left(2 \ip + p^2 +2 \up \right)(2 \phi + \delta \omega) = 4\pi n\omega.
	    \label{eq:GeneralSpiral}
	\end{equation}
	This can be re-arranged to give the equation of a spiral
	\begin{equation}
	    p=\pm \sqrt{
	      \frac{4 \pi n \omega}{2\phi + \delta \omega}
	      - 2\up -2\ip
	    }.
	    \label{eq:VortexCondition}
	\end{equation}
	From \eqref{eq:VortexCondition} for any particular angle $\phi$ there is a minimum value of $n$, below which there will be no real solutions for the radius $p$. 
	This is reminiscent of similar interference conditions that may be derived in the SFA e.g.~to describe above-threshold ionization (ATI) peaks \cite{Lewenstein1995}.
	The condition \eqref{eq:VortexCondition} is capable of describing a varying number of spiral arms.
	In \figref{fig:SpiralCond} we plot the spiral for different values of $n$ using both the positive and negative branches. In panels (a), (b) and (c) we use the first $7$, next  $8$ and the following $9$ integer values of $n$, respectively and both branches of \eqref{eq:VortexCondition}. Using more and higher values of $n$ leads the spiral arms filling a larger region, with some arms becoming crowded in the center, more evenly spaced for higher $p$ and finally developing gaps for the highest $p$. Thus, this suggests that in specific regions there will be a particular number of evenly spaced spiral arms (similar to Fermat spirals), which increases with $p$. In short, we expect the interference vortices to have an increasing number of spiral arms with $p$.
	
	\subsection{Computation of interference vortices}
	\label{Sec:MomentumDistribusions}
	\begin{figure*}
		\includegraphics[width=\linewidth]{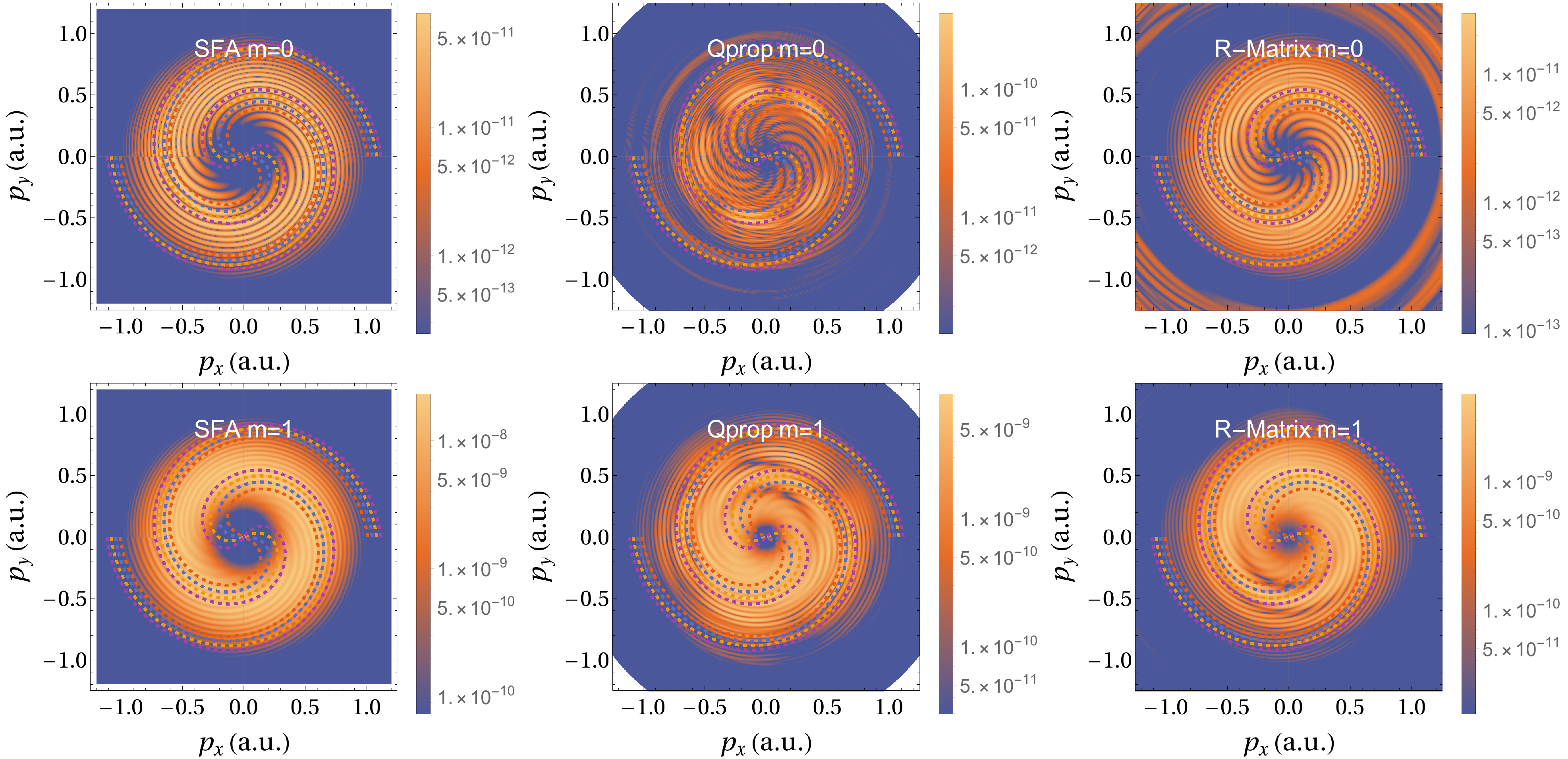}
		\caption{Momentum dependent photoelectron yield of $2p_0$ [top row] and $2p_1$ [bottom row] electrons ionized by two counter-rotating, 400~nm, $2.5\times 10^{13}$~W/cm$^2$,  4-cycle sin$^2$ pulses separated by a delay of 4 laser cycles, computed using the SFA [left], Qprop [centre] and the RMT method [right]. All figures show the vortex interferences, upon which we have superimposed the spiral condition derived above with dashed lines. The colour of the lines correspond to value of $n$ used in \eqref{eq:GeneralSpiral}. The Qprop and RMT momentum distributions sample states with $l>3$ in the spherical harmonic expansion to resolve the processes involving more than 3 photons. All panels are plotted on a logarithmic scale. A different scale is used for each panel to highlight the suppression of the photoelectron yield from $2p_0$ in comparison to $2p_1$. The units of the scale are arbitrary but comparisons can be made between computations using same model. The distributions are calculated in the $p_z = 0.1$ a.u.\;plane. This was chosen as at $p_z=0$, there is a node for $m=0$ leading to nearly zero signal. 
		}
		\label{fig:Vortices}
	\end{figure*}
	In this section we present numerical calculations of strong field ionization of neon via the two counter-rotating fields described above.
	In \figref{fig:Vortices} we plot the result of performing this calculation with the SFA [left column], Qprop [central column] and RMT [right column], for initial $2p$ orbitals of ground state neon with magnetic quantum number $m=0$ [top row] and $m=1$ [bottom row]. The vortices are clearly visible in all plots. The interference vortex condition is superimposed on all plots and matches very well, clearly capturing the core behaviour. It is notable that such a simplified SFA model is able to capture the features of the vortex. There are however differences, such as the hole in the center of the SFA results, which is less visible in the RMT and Qprop outcomes. This is likely due to the Coulomb force, which will decelerate the outgoing electron wavepacket and thus reduce its final momentum closing the hole. 
	This can be roughly accounted for in the SFA if the radial final momentum is taken to be lower than the initial, which will reduce the size of the hole.
	One clear aberration is the discontinuity along $\phi=0$ in the SFA results. In the SFA $\phi$ is fixed between $\pm\pi$ yielding a discontinuity. In the current construction the interference vortices can only be reproduced properly if $\phi$ can vary over a range larger than $2\pi$ in some regions, as it does in the interference condition. 
	
	In the RMT and Qprop momentum distribution calculation the spherical harmonic expansion was limited to to angular momenta $l>3$. The rationale for this is that the spiral interference arises from an ionization process involving at least 7 photons, which typically leads to population of states close to $l=7$. Population in states of low angular momenta are likely due to processes involving intermediate excited states, which are not accounted for by the SFA treatment. At the specific field parameters used here (peak intensity of $2.5\times 10^{13}$~W/cm$^2$ and 4-cycle sin$^2$ pulses), some contributions from such intermediate excited states can be identified for neon. This leads to additional interference vortices [not shown] superimposed on those presented in \figref{fig:Vortices}.
	   
	We find that the intermediate state effects are sensitive to the laser pulse parameters, and their contribution may be reduced if alternative parameters are used. We have verified that for higher intensities ($10^{14}$~W/cm$^2$) or long pulses (10-cycle sin$^2$ pulses), the low values of angular momenta do not play a significant role, and the interference vortices arising from at least 7-photon ionization from the ground state are the dominant effect. In fact very similar spirals can be produced with QProp and RMT using all angular momenta for an intensity of $10^{14}$~W/cm$^2$. This is aided by the fact the spiral fringes do not vary considerably with intensity in this regime. For longer pulses, the interference vortices are visible along with the characteristic ATI rings/ peaks, with the number of arms of each spiral increasing with the order of the ATI processes. A similar effect was observed in lower-order ATI processes in recent experiments \cite{Pengel2017}.

	Although it is difficult to resolve by eye, the number of spiral arms does vary, in \figref{fig:Vortices}, increasing with momentum. We have verified this by performing Fourier analysis on the momentum distributions across rings of fixed cylindrical radial momentum $p_{\perp}$ so that we can count the number of spiral arms intersecting each ring. This is also corroborated by the very good match with the interference condition. Thus, very good agreement can be achieved with a relatively simple SFA model, where many features can be identified. However, if we now examine the OAM of the outgoing  photoelectron we can achieve additional insight and understanding.

	\section{Interpretation of vortex in terms of OAM}
	\label{Sec:OAM}
	
	\subsection{Numerical OAM results}
	\label{Sec:OAMDistrbusions}
	\begin{figure*}
		\includegraphics[width=0.75\linewidth]{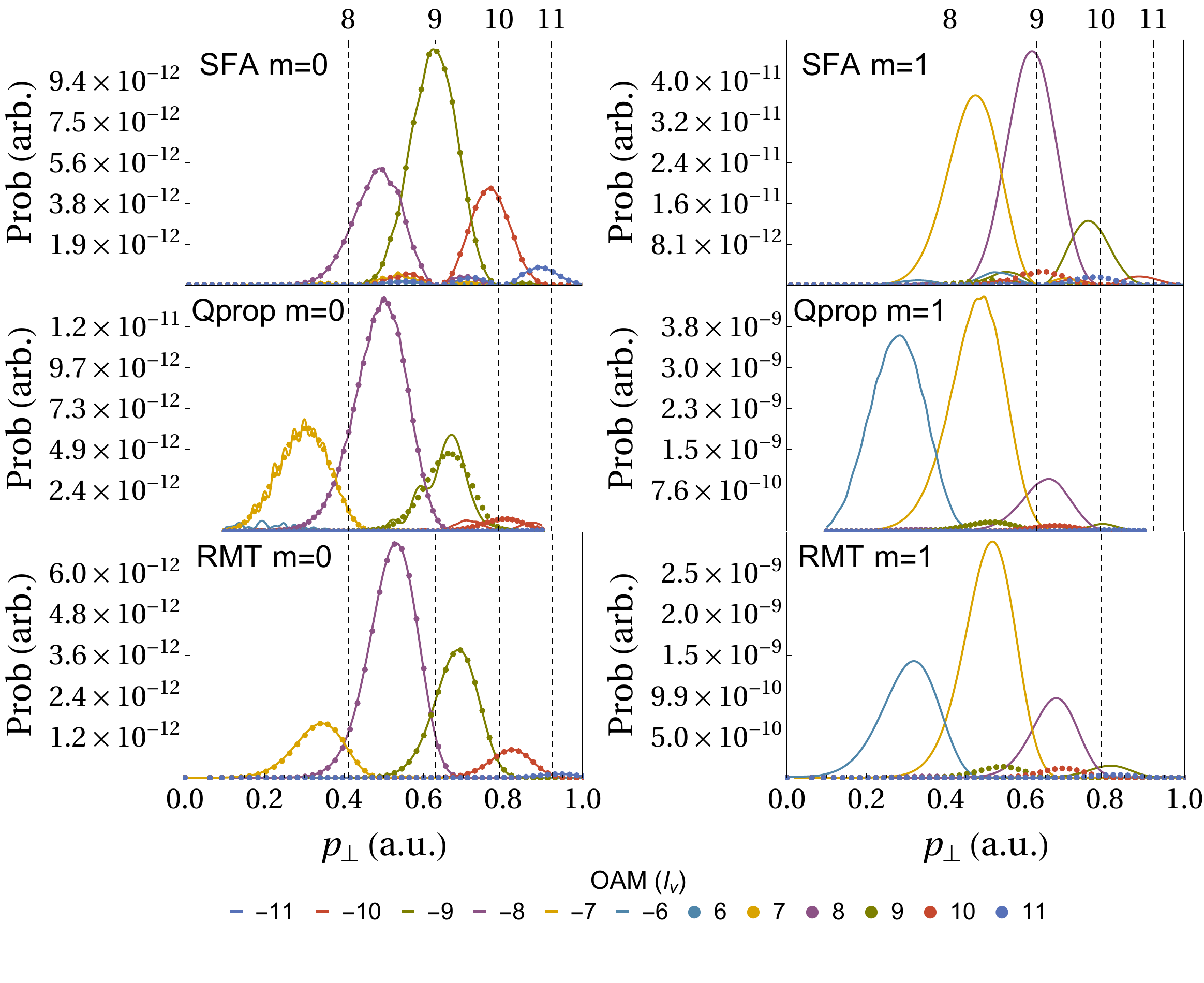}
		\caption{Photoelectron yield $\vert M_{l_v}(p_{\perp},p_{||})\vert^2$ for differing orbital angular momenta $l_v$. Computed using the SFA [top], Qprop [middle] and the RMT method [bottom]. The same field and target parameters have been used as in \figref{fig:Vortices}. The vertical dashed lines denote OAM regions from solving \eqref{eq:semiclassical}. 
		The left  and right columns shows computations for a $2p_0$ and $2p_1$ initial state, respectively.
		For positive values of $l_v$ the distribution is marked by data points, while for negative values a line plot is used. This is so that positive and negative OAM can be distinguished as they completely overlap in the $m=0$ case. 
		In the case of $m=1$ a marked asymmetry is apparent, with states of negative OAM dominating. This reflects the strong-field preference of the $m=1$ electron to be ionized by a pulse of opposite (negative) helicity, which drives transitions to negative OAM values.
		As in \figref{fig:Vortices} the OAM distributions were computed for $p_z=0.1$ a.u.
		}
			\label{fig:OAMDistribusion2D}
	\end{figure*}
	Using \eqref{eq:OAMTransformation} and \eqref{eq:OAMSphericalTransformation} we compute the `OAM distributions' for the three models presented.
	The \emph{vortex state} basis enables a description of the interference vortices in terms of the OAM of the freed electron. In \figref{fig:OAMDistribusion2D} we show the distribution of OAM over perpendicular momentum $p_{\perp}=\sqrt{p_x^2+p_y^2}$ for a fixed $p_z=0.1$~a.u., chosen as there is a node at $p_z=0$ leading to very low ionization signal for the case $m=0$. The left column of \figref{fig:OAMDistribusion2D} displays results for neon initially in the $m=0$, $p$-state. All distributions show a single peak corresponding to each of three momentum regions, with a small, fourth peak at the highest momentum shown. For each peak there are actually two overlapping sets of lines corresponding to OAM values with opposite signs (negative OAM is shown by a solid line while positive OAM is dotted), and it is the interference between these that leads to the interference vortices.  The RMT and Qprop OAM distributions are very similar but they differ somewhat from the SFA. In particular, there is a slight shift in the peak positions and relative heights. The highest peak in the SFA is for $l_v=\pm 9$, while it is $l_v=\pm8$ for RMT and Qprop. This is related to the larger `hole' in the SFA results (\figref{fig:Vortices}) and most likely can be traced to Coulomb distortions. There is also a shift of around half a photon's energy between the SFA peaks and those from RMT and Qprop.	On each panel in \figref{fig:OAMDistribusion2D} the ATI peaks are marked by vertical dashed lines, this can be interpreted as the number of photons that must be absorbed to reach a particular momentum. Despite the shift of the peaks between the models, in all cases the order of the ATI peak corresponds approximately to the OAM gained. 
	Thus we are extending the interpretation of the multiphoton picture, where each additional photon gained can add $\pm1$ to the OAM.
	
	On the right-hand side of \figref{fig:OAMDistribusion2D} the results for the $m=1$, $p$-state of neon are shown. The main effect (relative to the $m=0$ case) is the suppression of the positive OAM values, which is what causes the blurring for the equivalent results in \figref{fig:Vortices}. This asymmetry reflects the preferential ionization of $m=1$ electrons by a field of negative helicity. Such a field will tend to decrease the OAM of the ejected electron, and populate states of negative OAM. The asymmetry is a strong-field effect, quite different to the symmetric yields observed in previous studies of few-photon ionization of $s$ electrons. We see in \figref{fig:OAMDistribusion2D} that in this case states with OAM values symmetric about $l_v=1$ contribute over a common range of momenta. In particular, both Qprop and RMT calculations find that the peaks for $l_v=-7$ and $l_v=9$ are now aligned at $p=0.5$~a.u. (and similarly $l_v=-8$ and $l_v=10$ contribute at around $p=0.65$~a.u.).
	Despite their asymmetric yields, their interference is still sufficient to give rise to the spiral interference pattern observed in \figref{fig:Vortices}. 
	Thus, it is clear that at a given momentum, the photoelectron yield is contained in states whose OAM value is shifted by $+1$ relative to the $m=0$ case, suggesting that the magnetic quantum number is simply added to the final OAM. 
    For $m=-1$ electrons (not shown), the opposite tendency appears, and states of positive OAM dominate.

	\subsection{Building interference vortices from twisted electrons}
	\label{Sec:OAMSpiral}
	\begin{figure*}
		\includegraphics[width=\linewidth]{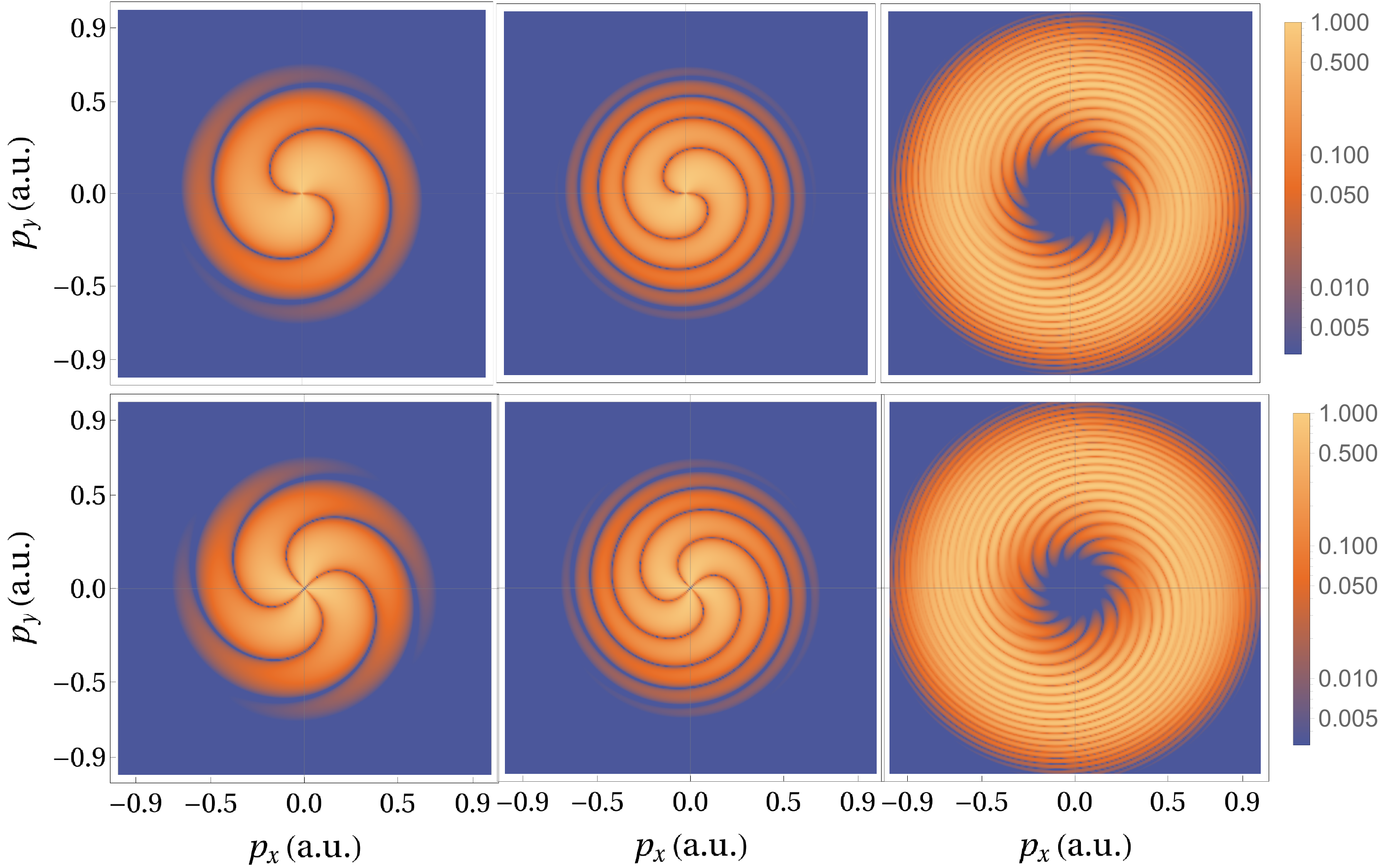}
		\caption{ Interference vortices from twisted electrons. Panels (a) and (b) show interference between photoelectrons with an OAM of $\pm 1$, panels (d) and (e) shows interference between photoelectrons with an OAM of $\pm 2$, the left column uses a delay of 4 cycles the middle column uses a delay of 1 cycle. The distribution are chosen to be Gaussian in momentum coordinates $p_x$ and $p_y$. The right column shows the interference if the OAM distributions of the SFA [top] and RMT [bottom]. Gaussian distribution of twisted electron states are taken to reflect the OAM distributions show in \figref{fig:OAMDistribusion2D}. Each panel is normalized with respect to its peak value and plotted on a logarithmic scale.}
		\label{fig:OAMReconstruction}	
	\end{figure*}
	
	Using the distributions presented in \figref{fig:OAMDistribusion2D} the interference vortices can be reconstructed. This provides particular insight into the meaning of the OAM in a strong field context. To recreate this interference, for the case of $m=0$, we will consider a distribution of outgoing photoelectrons with two opposite values of OAM $\pm l_v$. For an initial state of general $m$ one should consider two states with OAM $m\pm l_v$. This can be written as
	\begin{equation}
	    \ket{\psi(t)}=\int d^2 \pb' w(\pb')\left( 
	        \ket{\psi_{m+l_v, \pb'}(t)} + \ket{\psi_{m-l_v, \pb'}(t)}
	    \right).
	    \label{eq:BuildSpiral}
	\end{equation}
	Here $ \ket{\psi_{m\pm l_v, \pb}(t)}$ is electron vortex state with OAM $m\pm l_v$ and momentum $\pb=(p_{||},p_{\perp})$ and $w(\pb)$ provides a weighting over the cylindrical momentum coordinates. 
	Now, if we project this state onto a 3D plane wave momentum state we can examine the interference in the $p_x p_y$-plane,
	\begin{align}
	    \braket{\pb|\psi(t)}=
	    \frac{i^{-m}}{\pi} w(\pb) e^{-i/4 p^2 t}
	    \cos\left( \phi l_{v} + \frac{1}{4}p^2 \delta -\frac{\pi l_v}{2}\right).
	\end{align}
	The $m$ cancels in the interference fringes, as they depend on the difference of the values of OAM.
	We have used the following momentum representation of Bessel electron vortex states \cite{Lloyd2017}
	\begin{equation}
	    \braket{\pb|\psi_{l_v \pb'}( t)}=
	    \frac{i^{-l_v}e^{i l_v \phi}e^{-i/2 p^2 t}}{2\pi p_{\perp}}
	    \delta(p_{||}' - p_{||}) \delta(p_{\perp}' -p_{\perp}).
	\end{equation}
	Thus, we identify the interference between the two vortex states as leading to the condition
	\begin{equation}
	    4\phi l_v + \delta p^2 =2 \pi( 2n+l_v).
	    \label{eq:Fermat}
	\end{equation}
	This result generalizes the expressions derived in Refs.\;\cite{NgokoDjiokap2015, NgokoDjiokap2016} for one-photon and two-photon ionization to the strong field regime, and describes a Fermat spiral with $2 l_v$ arms. In \figref{fig:OAMReconstruction} [(a-b) and (d-e)], we plot examples $l_v=1$ and $l_v=2$, for different values of $\delta$, which inversely sets the growth of the radius with respect to the angle $\phi$. These spirals are reminiscent of those presented in Refs.\;\cite{NgokoDjiokap2015,NgokoDjiokap2016}, and observe a similar delay dependence.
	In panels (c) and (f) we plot a reconstruction of the full interference vortices by combining pairs of OAM in different momentum regions as indicated by the distributions given in \figref{fig:OAMDistribusion2D}. The SFA  and RMT distributions for $m=0$ are recreated in panels (c) and (f), respectively. This interference construction will also hold for $m=1$, the only difference being positive OAM are suppressed. This explains why the original spiral condition works so well for both cases.
	For each pair of OAM values included the weighting function $w(\pb)$ is set as a Gaussian over $p_{\perp}$ to reflect the position, heights and widths of the respective OAM distributions.
	
	This clearly demonstrates the link between interference vortices and the OAM of the photoelectrons. However, we have two separate spiral conditions. From the interference of photoelectrons of opposite OAM, \eqref{eq:Fermat} leads to a condition for Fermat spirals with a fixed number of arms determined by the OAM. But due to the localization of OAM of the photoelectron, this condition is only valid in specific momentum regions and the number of arms of the interference vortex does not remain fixed. Alternatively, from the SFA, \eqref{eq:GeneralSpiral}, we derived a condition capable of describing a varying number of arms. It is, however, possible to reconcile these two, seemingly disparate descriptions. By employing the semiclassical relation
	\begin{equation}
	    2\omega l_v(p):=p^2+2\ip+2\up,
	    \label{eq:semiclassical}
	\end{equation}
	in \eqref{eq:GeneralSpiral} we recover the same form as \eqref{eq:Fermat}, 
	\begin{align}
	    4\phi l_v(p) +\delta p^2 = 4\pi \left(n-N_c \; l_v(0)\;\right),
	    \label{eq:CombinedCondition}
	\end{align}
	where $\delta=2\pi N_c/\omega$ is the number of laser cycles in the delay. Thus, there is a $p$-dependent semiclassical expression for the OAM, $l_v(p)$, which maps directly onto the ATI equation and can be interpreted as each additional photon contributing $\pm1$ to the OAM.
	The integer $n$ is shifted by $N_c \times l(0)$, where $l(0)$ corresponds to lowest possible OAM given $l(0)$ photons are required to overcome the barrier $\ip+\up$.
	The expression given by \eqref{eq:CombinedCondition} maps the SFA condition onto vortex Fermat spirals. It demonstrates, via this semiclassical relation, in specific regions where $l(p)$ is an integer, that \eqref{eq:CombinedCondition} will behave like the condition given by \eqref{eq:Fermat}.

	\section{Discussion and Conclusions}
	\label{Sec:Conclusions}
	
	In the previous sections the main dynamics of the interference vortices in the strong field regime were captured and new light was shed on their formation by analysing the orbital angular momenta (OAM). In this section we will place the interference vortices and photoelectron OAM in a wider context, making parallels with other systems as well as discussing the potential to measure the OAM and exploit it in time-resolved imaging. 
	
	Vortices and spirals in a physical system are indicative of rotational dynamics and symmetries.
    When we talk about vortices in quantum matter the physics of liquids is the first example that comes to mind \cite{Fetter09}. In the present  context, however, the analogy to Bose Einstein condensates (BEC) of dilute atomic gases is more appropriate  \cite{PitaevskiiStringari2016book}. Vortices in BEC can be created by rotations (like in a bucket of liquid Helium)
    \cite{Madison00}, that is ``stirring with an optical spoon'' \cite{Srinivasan06}  or a ``phase imprinting'' \cite{Dobrek1999}.  While the vortices of charges higher than $\pm 1$ are not stable in these systems, rotating BECs form beautiful Abrikosov lattices of charge one vortices. Still, interference  of matter waves was, from the very beginning, proposed as a way  to observe topological defects \cite{Bolda98,Tempere98,Dalfovo99}, stimulated by non-linear optics \cite{Staliunas99,Basistiy93,Kreminskaya98,Nye74}. Interference of a vortex state with a plain BEC 
    leads to the famous fork patterns \cite{Fetter09, PitaevskiiStringari2016book}. The interference of two BECs with vortices leads to spiral patterns similar to those discussed here \cite{Fetter09, PitaevskiiStringari2016book}, except that i) the patterns in BEC are formed in the density profile in real space and ii) they are formed by a macroscopic quantum  state, and are thus robust with respect to decoherence.

    The question of how to measure the OAM of photoelectrons in strong field ionization may also be answered via interference.
    The main difference here is that interference structures are formed in the electron momentum space.  
    The interference vortices are in principle measurable in experiments with angular resolution of the electrons created in ionization. Indeed, in the multi-photon regime the experimental measurements clearly showed the Fermat spirals \cite{Pengel2017,Pengel2017a}.
    Still, the problem is how sensitive the interference patterns may be to the parameters of the system. Will incoherent effects in the strong field regime blur the interference? Preliminary simulations of focal averaging in  our regime of strong and  non-perturbative laser fields, indicate that the feasibility of observing the interference pattern is robust. The interference fringes do not vary considerably with intensity until $\up>\ip$, which for neon at $\lambda=400$~nm occurs at intensities greater than $7\times 10^{14}$~W/cm$^2$.
    Going to higher intensities or longer wavelength would perhaps require the use of pre-designed laser pulses with a flat top, for instance \cite{Boutu2011,Toma1999}, using more sophisticated methods.
    It is also important that the strong field ionization leading to the interference vortices is the dominant process. For lower intensities and multielectron targets, coupling to excited states can occur. Among other effects this could cause superposed interference vortices from both the ground and excited states and the combination can be difficult to disentangle.
    For larger intensities and longer pulses these effects can be minimized. For more complex targets such as molecules contributions from multiple states or molecular centres may also present difficulties as again the superposition of many interference vortices, associated with different ionization potentials and magnetic quantum numbers, could make them difficult to resolve.

    For the case of circular fields we have presented a clear path to determine the OAM of the photoelectrons, via a second field of opposite helicity. What about elliptical, linear or more complex fields? In the case of linear fields, as they do not interact with the OAM, the same scheme as discussed in this work would not be possible. It would, however, for any field be possible to produce a well-characterized reference for measurement of OAM via interference using a circular pulse. Interference of pairs of twisted electron states will always yield a spiral in the $p_x p_y$-plane with the number of arms being equal to the difference between the two OAM values. Thus, a careful choices of circular pulses following any target pulse (after a delay) could reveal the OAM distribution. Exploiting the localisation of OAM  in momentum space for a circular field would lead to different interference vortices in different regions, in effect `scanning' the OAM of the photoelectrons ionized from the target pulse.
    However, this method would start to break down if multiple OAMs populate the same region of momentum space for a particular target pulse. Multiple vortices may be difficult to resolve and lead to interferences too complex to analyse.

	Another approach is to measure the OAM of the outgoing photoelectrons directly. This would require significant alteration to standard strong field experimental setups, such as velocity map imaging (VMI) \cite{Eppink1997, Takahashi2000} or a reaction microscope (ReMi) detection system \cite{Moshammer1996, Dorner2000, Ullrich2003}, so needs some justification. However, it would not only be a much more generalized and robust way to determine the OAM. Direct measurement could also allow for measurement of incompatible observables, which is not obviously possible using interferometric schemes via tailored laser fields. Incompatible observables are exploited in various aspects of quantum information and metrology including quantum key distribution, Bell inequalities and quantum sensing \cite{Heinosaari2016, Bongs2019}. Additionally, the measurement of OAM could enhance existing strong field procedures. For example, it would allow direct measurement of the quantum magnetic number $m$ of the initial state.
	Furthermore, it has already be shown that recolliding twisted photoelectrons are sensitive to chiral molecular targets \cite{Tolstikhin2019}, which may be exploited for imaging. This may also have implications for photoelectron holography \cite{faria_it_2020}, where recollision/ reinteraction with the target plays an important role and chiral phases could be revealed. The inclusion of fields with high ellipticity and recollision probability, such as bi-circular fields, would enable manipulation of the OAM while retaining strong interaction with the target.
	Thus, OAM measurements could open up a range of possibilities for strong field research. 
	
	So what is the state of the art in OAM measurement and how can this be applied to strong field systems?
	In \cite{Bliokh2017,Lloyd2017} a review of state of measurement in OAM is given. A range of methodologies are available, an example of a typical approach is using diffraction via holograms or particularly shaped apertures, converting the phase information into spatial information. The first/most well-known method uses a fork phase mask \cite{Guzzinati2014, Saitoh2013} but there are difficulties such as the the limited transparency of the mask. In \cite{Grillo2017} the OAM sorter developed transforms the OAM into spatial position, reminiscent of a Stern-Gerlach style measurement of spin. Such a scheme could be envisaged in a strong field set-up. This would forgo the usual momentum information from the $p_x p_y$-plane in favour of a $p_{\perp}$ and $l$ measurement. In theory this could be achieved in existing detection systems with the addition the hologram/phase mask and corrector as described in the OAM sorter \cite{Grillo2017}. This would map one of the dimensions in the $p_x p_y$-plane to the OAM, leading to different spots along the detector for different values of OAM.
	However, there are a great many practical considerations including the kinetic energy of the photoelectrons, propagation distance, detector and mask efficiencies to name a small few. In trying to combine these systems together it may be that they are incompatible without major alterations. However, the additional control and information that may be gleaned from strong field systems as a result certainly makes it worth consideration. 
	
	Because the OAM gives a direct measurement of the quantum magnetic number as well as providing sensitivity to chiral molecules \cite{Asenjo-Garcia2014, Tolstikhin2019}, it is a useful tool for photo-induced time-resolved measurements.
	The first thought may be in exploiting OAM for such measurements through the incident light such as in twisted attosecond pulses. This is a non-trivial task but progress has been made in high-harmonic generation table top sources \cite{Dorney2019, Wang2019} as well in free-electron lasers \cite{Rebernik2017}. However, the focus of this work is on the OAM of the outgoing photoelectron, which may be exploited via detection. As such there would be many possibilities for pump-probe schemes. For example the two counter-rotating circular fields employed in this work could each act as a probe, while two pumps separated by the same delay could be i)	XUV attosecond pulses or ii) few-cycle IR fields. In the case of i) we have a situation reminiscent of the RABBITT (reconstruction of attosecond harmonic beating by interference of two-photon transitions) and attosecond streaking techniques \cite{Vrakking2014,Cattaneo2016}. In both cases, i) and ii), the idea would be to use VMI to exploit the interference, counting the number of spirals to determine the OAM. 
	In order to apply such methods the interference vortices should be further studied in more complex targets to address the above-mentioned reservations for molecules.
	Direct measurement of the OAM could allow for any pump probe configuration as well as potentially removing the difficulties of using more complex targets. The ability to employ a variety pump-probe scheme with any ellipticity could allow for non-trivial coupling of the laser pulses to the OAM and novel time-resolved measurements of photo-induced dynamics.
	
	In conclusion, we have presented a description of interference vortices in the strong field regime. Using the strong field approximation we have captured the main physical mechanism, deriving an interference condition, which closely matches calculations performed with the time-dependent Schr\"odinger equation-based solvers Qprop and R-Matrix with time dependence. 
	We can explain the blurring of interference in the $m=1$ case by the asymmetric yield from circular fields of opposite helicity, which is an exclusively strong field effect.
	By examining the system using orbital angular momentum (OAM) of twisted photoelectron states, we find a new interpretation of the electron vortices in terms of the interference of pairs of OAM states. Not only do we find good agreement between all models here but we also uncover a semiclassical relationship for the OAM by considering the interference between pairs of vortex states and linking it to the condition derived using the SFA. This has consequences, opening up the possibility to measure the OAM directly, or indirectly using interferometric schemes. The OAM of photoelectrons can open the possibility to a range of novel measurements and control in strong field systems.
	
	\section*{Conflicts of interest}
	There are no conflicts of interest to declare.

    \section*{Acknowledgements}
    This work was in part funded by the UK Engineering and Physical Sciences Research Council (EPSRC). ASM acknowledges grant EP/P510270/1, which is within the remit of the InQuBATE Skills Hub for Quantum Systems Engineering. CFMF would like to acknowledge EPSRC grant EP/T019530/1. GSJA and ACB acknowledge EPSRC grants EP/P022146/1, EP/P013953/1 and EP/R029342/1.
    The RMT code is part of the UK-AMOR suite, and can be obtained for free at Ref.~\cite{repo}. This work benefited from computational support by CoSeC, the Computational Science Centre for Research Communities, through CCPQ.  This work used the ARCHER UK National Supercomputing Service (\url{www.archer.ac.uk}), for which access was obtained via the UK-AMOR consortium funded by EPSRC.
    
    ICFO group acknowledges support from ERC AdG NOQIA, Spanish Ministry of Economy and Competitiveness (``Severo Ochoa'' program for Centres of Excellence in R\&D (CEX2019-000910-S), Plan National FISICATEAMO and FIDEUA PID2019-106901GB-I00/10.13039 / 501100011033, FPI), Fundació Privada Cellex, Fundació Mir-Puig, and from Generalitat de Catalunya (AGAUR Grant No. 2017 SGR 1341, CERCA program, QuantumCAT \textunderscore U16-011424, co-funded by ERDF Operational Program of Catalonia 2014-2020), MINECO-EU QUANTERA MAQS (funded by State Research Agency (AEI) PCI2019-111828-2 / 10.13039/501100011033), EU Horizon 2020 FET-OPEN OPTOLogic (Grant No 899794), and the National Science Centre, Poland-Symfonia Grant No. 2016/20/W/ST4/00314.

	\bibliography{OAMinSF.bib}

\begin{thebibliography}{70}%
\makeatletter
\providecommand \@ifxundefined [1]{%
 \@ifx{#1\undefined}
}%
\providecommand \@ifnum [1]{%
 \ifnum #1\expandafter \@firstoftwo
 \else \expandafter \@secondoftwo
 \fi
}%
\providecommand \@ifx [1]{%
 \ifx #1\expandafter \@firstoftwo
 \else \expandafter \@secondoftwo
 \fi
}%
\providecommand \natexlab [1]{#1}%
\providecommand \enquote  [1]{``#1''}%
\providecommand \bibnamefont  [1]{#1}%
\providecommand \bibfnamefont [1]{#1}%
\providecommand \citenamefont [1]{#1}%
\providecommand \href@noop [0]{\@secondoftwo}%
\providecommand \href [0]{\begingroup \@sanitize@url \@href}%
\providecommand \@href[1]{\@@startlink{#1}\@@href}%
\providecommand \@@href[1]{\endgroup#1\@@endlink}%
\providecommand \@sanitize@url [0]{\catcode `\\12\catcode `\$12\catcode
  `\&12\catcode `\#12\catcode `\^12\catcode `\_12\catcode `\%12\relax}%
\providecommand \@@startlink[1]{}%
\providecommand \@@endlink[0]{}%
\providecommand \url  [0]{\begingroup\@sanitize@url \@url }%
\providecommand \@url [1]{\endgroup\@href {#1}{\urlprefix }}%
\providecommand \urlprefix  [0]{URL }%
\providecommand \Eprint [0]{\href }%
\providecommand \doibase [0]{https://doi.org/}%
\providecommand \selectlanguage [0]{\@gobble}%
\providecommand \bibinfo  [0]{\@secondoftwo}%
\providecommand \bibfield  [0]{\@secondoftwo}%
\providecommand \translation [1]{[#1]}%
\providecommand \BibitemOpen [0]{}%
\providecommand \bibitemStop [0]{}%
\providecommand \bibitemNoStop [0]{.\EOS\space}%
\providecommand \EOS [0]{\spacefactor3000\relax}%
\providecommand \BibitemShut  [1]{\csname bibitem#1\endcsname}%
\let\auto@bib@innerbib\@empty
\bibitem [{\citenamefont {Allen}\ \emph {et~al.}(1992)\citenamefont {Allen},
  \citenamefont {Beijersbergen}, \citenamefont {Spreeuw},\ and\ \citenamefont
  {Woerdman}}]{Allen1992}%
  \BibitemOpen
  \bibfield  {author} {\bibinfo {author} {\bibfnamefont {L.}~\bibnamefont
  {Allen}}, \bibinfo {author} {\bibfnamefont {M.~W.}\ \bibnamefont
  {Beijersbergen}}, \bibinfo {author} {\bibfnamefont {R.~J.~C.}\ \bibnamefont
  {Spreeuw}},\ and\ \bibinfo {author} {\bibfnamefont {J.~P.}\ \bibnamefont
  {Woerdman}},\ }\href {https://doi.org/10.1103/PhysRevA.45.8185} {\bibfield
  {journal} {\bibinfo  {journal} {Phys. Rev. A}\ }\textbf {\bibinfo {volume}
  {45}},\ \bibinfo {pages} {8185} (\bibinfo {year} {1992})}\BibitemShut
  {NoStop}%
\bibitem [{\citenamefont {Bliokh}\ \emph {et~al.}(2007)\citenamefont {Bliokh},
  \citenamefont {Bliokh}, \citenamefont {Savel'ev},\ and\ \citenamefont
  {Nori}}]{Bliokh2007}%
  \BibitemOpen
  \bibfield  {author} {\bibinfo {author} {\bibfnamefont {K.~Y.}\ \bibnamefont
  {Bliokh}}, \bibinfo {author} {\bibfnamefont {Y.~P.}\ \bibnamefont {Bliokh}},
  \bibinfo {author} {\bibfnamefont {S.}~\bibnamefont {Savel'ev}},\ and\
  \bibinfo {author} {\bibfnamefont {F.}~\bibnamefont {Nori}},\ }\href
  {https://doi.org/10.1103/PhysRevLett.99.190404} {\bibfield  {journal}
  {\bibinfo  {journal} {Phys. Rev. Lett.}\ }\textbf {\bibinfo {volume} {99}},\
  \bibinfo {pages} {190404} (\bibinfo {year} {2007})}\BibitemShut {NoStop}%
\bibitem [{\citenamefont {Bliokh}\ \emph {et~al.}(2017)\citenamefont {Bliokh},
  \citenamefont {Ivanov}, \citenamefont {Guzzinati}, \citenamefont {Clark},
  \citenamefont {{Van Boxem}}, \citenamefont {B{\'{e}}ch{\'{e}}}, \citenamefont
  {Juchtmans}, \citenamefont {Alonso}, \citenamefont {Schattschneider},
  \citenamefont {Nori},\ and\ \citenamefont {Verbeeck}}]{Bliokh2017}%
  \BibitemOpen
  \bibfield  {author} {\bibinfo {author} {\bibfnamefont {K.~Y.}\ \bibnamefont
  {Bliokh}}, \bibinfo {author} {\bibfnamefont {I.~P.}\ \bibnamefont {Ivanov}},
  \bibinfo {author} {\bibfnamefont {G.}~\bibnamefont {Guzzinati}}, \bibinfo
  {author} {\bibfnamefont {L.}~\bibnamefont {Clark}}, \bibinfo {author}
  {\bibfnamefont {R.}~\bibnamefont {{Van Boxem}}}, \bibinfo {author}
  {\bibfnamefont {A.}~\bibnamefont {B{\'{e}}ch{\'{e}}}}, \bibinfo {author}
  {\bibfnamefont {R.}~\bibnamefont {Juchtmans}}, \bibinfo {author}
  {\bibfnamefont {M.~A.}\ \bibnamefont {Alonso}}, \bibinfo {author}
  {\bibfnamefont {P.}~\bibnamefont {Schattschneider}}, \bibinfo {author}
  {\bibfnamefont {F.}~\bibnamefont {Nori}},\ and\ \bibinfo {author}
  {\bibfnamefont {J.}~\bibnamefont {Verbeeck}},\ }\href
  {https://doi.org/10.1016/j.physrep.2017.05.006} {\bibfield  {journal}
  {\bibinfo  {journal} {Phys. Rep.}\ }\textbf {\bibinfo {volume} {690}},\
  \bibinfo {pages} {1} (\bibinfo {year} {2017})}\BibitemShut {NoStop}%
\bibitem [{\citenamefont {Lloyd}\ \emph {et~al.}(2017)\citenamefont {Lloyd},
  \citenamefont {Babiker}, \citenamefont {Thirunavukkarasu},\ and\
  \citenamefont {Yuan}}]{Lloyd2017}%
  \BibitemOpen
  \bibfield  {author} {\bibinfo {author} {\bibfnamefont {S.~M.}\ \bibnamefont
  {Lloyd}}, \bibinfo {author} {\bibfnamefont {M.}~\bibnamefont {Babiker}},
  \bibinfo {author} {\bibfnamefont {G.}~\bibnamefont {Thirunavukkarasu}},\ and\
  \bibinfo {author} {\bibfnamefont {J.}~\bibnamefont {Yuan}},\ }\href
  {https://doi.org/10.1103/RevModPhys.89.035004} {\bibfield  {journal}
  {\bibinfo  {journal} {Rev. Mod. Phys.}\ }\textbf {\bibinfo {volume} {89}},\
  \bibinfo {pages} {035004} (\bibinfo {year} {2017})}\BibitemShut {NoStop}%
\bibitem [{\citenamefont {Z{\"u}rch}\ \emph {et~al.}(2012)\citenamefont
  {Z{\"u}rch}, \citenamefont {Kern}, \citenamefont {Hansinger}, \citenamefont
  {Dreischuh},\ and\ \citenamefont {Spielmann}}]{Zurch2012}%
  \BibitemOpen
  \bibfield  {author} {\bibinfo {author} {\bibfnamefont {M.}~\bibnamefont
  {Z{\"u}rch}}, \bibinfo {author} {\bibfnamefont {C.}~\bibnamefont {Kern}},
  \bibinfo {author} {\bibfnamefont {P.}~\bibnamefont {Hansinger}}, \bibinfo
  {author} {\bibfnamefont {A.}~\bibnamefont {Dreischuh}},\ and\ \bibinfo
  {author} {\bibfnamefont {C.}~\bibnamefont {Spielmann}},\ }\href
  {https://doi.org/10.1038/nphys2397} {\bibfield  {journal} {\bibinfo
  {journal} {Nat. Phys.}\ }\textbf {\bibinfo {volume} {8}},\ \bibinfo {pages}
  {743} (\bibinfo {year} {2012})}\BibitemShut {NoStop}%
\bibitem [{\citenamefont {Rego}\ \emph {et~al.}(2019)\citenamefont {Rego},
  \citenamefont {Dorney}, \citenamefont {Brooks}, \citenamefont {Nguyen},
  \citenamefont {Liao}, \citenamefont {Rom{\'{a}}n}, \citenamefont {Couch},
  \citenamefont {Liu}, \citenamefont {Pisanty}, \citenamefont {Lewenstein},
  \citenamefont {Plaja}, \citenamefont {Kapteyn}, \citenamefont {Murnane},\
  and\ \citenamefont {Hern{\'{a}}ndez-Garc{\'{i}}a}}]{Rego2019}%
  \BibitemOpen
  \bibfield  {author} {\bibinfo {author} {\bibfnamefont {L.}~\bibnamefont
  {Rego}}, \bibinfo {author} {\bibfnamefont {K.~M.}\ \bibnamefont {Dorney}},
  \bibinfo {author} {\bibfnamefont {N.~J.}\ \bibnamefont {Brooks}}, \bibinfo
  {author} {\bibfnamefont {Q.~L.}\ \bibnamefont {Nguyen}}, \bibinfo {author}
  {\bibfnamefont {C.~T.}\ \bibnamefont {Liao}}, \bibinfo {author}
  {\bibfnamefont {J.~S.}\ \bibnamefont {Rom{\'{a}}n}}, \bibinfo {author}
  {\bibfnamefont {D.~E.}\ \bibnamefont {Couch}}, \bibinfo {author}
  {\bibfnamefont {A.}~\bibnamefont {Liu}}, \bibinfo {author} {\bibfnamefont
  {E.}~\bibnamefont {Pisanty}}, \bibinfo {author} {\bibfnamefont
  {M.}~\bibnamefont {Lewenstein}}, \bibinfo {author} {\bibfnamefont
  {L.}~\bibnamefont {Plaja}}, \bibinfo {author} {\bibfnamefont {H.~C.}\
  \bibnamefont {Kapteyn}}, \bibinfo {author} {\bibfnamefont {M.~M.}\
  \bibnamefont {Murnane}},\ and\ \bibinfo {author} {\bibfnamefont
  {C.}~\bibnamefont {Hern{\'{a}}ndez-Garc{\'{i}}a}},\ }\href
  {https://doi.org/10.1126/science.aaw9486} {\bibfield  {journal} {\bibinfo
  {journal} {Science}\ }\textbf {\bibinfo {volume} {364}},\ \bibinfo {pages}
  {1253} (\bibinfo {year} {2019})}\BibitemShut {NoStop}%
\bibitem [{\citenamefont {Pisanty}\ \emph {et~al.}(2019)\citenamefont
  {Pisanty}, \citenamefont {Machado}, \citenamefont
  {Vicu{\~{n}}a-Hern{\'{a}}ndez}, \citenamefont {Pic{\'{o}}n}, \citenamefont
  {Celi}, \citenamefont {Torres},\ and\ \citenamefont
  {Lewenstein}}]{Pisanty2019}%
  \BibitemOpen
  \bibfield  {author} {\bibinfo {author} {\bibfnamefont {E.}~\bibnamefont
  {Pisanty}}, \bibinfo {author} {\bibfnamefont {G.~J.}\ \bibnamefont
  {Machado}}, \bibinfo {author} {\bibfnamefont {V.}~\bibnamefont
  {Vicu{\~{n}}a-Hern{\'{a}}ndez}}, \bibinfo {author} {\bibfnamefont
  {A.}~\bibnamefont {Pic{\'{o}}n}}, \bibinfo {author} {\bibfnamefont
  {A.}~\bibnamefont {Celi}}, \bibinfo {author} {\bibfnamefont {J.~P.}\
  \bibnamefont {Torres}},\ and\ \bibinfo {author} {\bibfnamefont
  {M.}~\bibnamefont {Lewenstein}},\ }\href
  {https://doi.org/10.1038/s41566-019-0450-2} {\bibfield  {journal} {\bibinfo
  {journal} {Nat. Photonics}\ }\textbf {\bibinfo {volume} {13}},\ \bibinfo
  {pages} {569} (\bibinfo {year} {2019})}\BibitemShut {NoStop}%
\bibitem [{\citenamefont {Gemsheim}\ and\ \citenamefont
  {Rost}(2019)}]{Gemsheim2019}%
  \BibitemOpen
  \bibfield  {author} {\bibinfo {author} {\bibfnamefont {S.}~\bibnamefont
  {Gemsheim}}\ and\ \bibinfo {author} {\bibfnamefont {J.-M.}\ \bibnamefont
  {Rost}},\ }\href {https://doi.org/10.1103/PhysRevA.100.043408} {\bibfield
  {journal} {\bibinfo  {journal} {Phys. Rev. A}\ }\textbf {\bibinfo {volume}
  {100}},\ \bibinfo {pages} {43408} (\bibinfo {year} {2019})},\ \Eprint
  {https://arxiv.org/abs/1909.00728} {arXiv:1909.00728} \BibitemShut {NoStop}%
\bibitem [{\citenamefont {V\'elez}\ \emph {et~al.}(2018)\citenamefont
  {V\'elez}, \citenamefont {Krajewska},\ and\ \citenamefont
  {Kami\ifmmode~\acute{n}\else \'{n}\fi{}ski}}]{Velez2018}%
  \BibitemOpen
  \bibfield  {author} {\bibinfo {author} {\bibfnamefont {F.~C.}\ \bibnamefont
  {V\'elez}}, \bibinfo {author} {\bibfnamefont {K.}~\bibnamefont {Krajewska}},\
  and\ \bibinfo {author} {\bibfnamefont {J.~Z.}\ \bibnamefont
  {Kami\ifmmode~\acute{n}\else \'{n}\fi{}ski}},\ }\href
  {https://doi.org/10.1103/PhysRevA.97.043421} {\bibfield  {journal} {\bibinfo
  {journal} {Phys. Rev. A}\ }\textbf {\bibinfo {volume} {97}},\ \bibinfo
  {pages} {043421} (\bibinfo {year} {2018})}\BibitemShut {NoStop}%
\bibitem [{\citenamefont {Cajiao~V\'elez}\ \emph {et~al.}(2020)\citenamefont
  {Cajiao~V\'elez}, \citenamefont {Kami\ifmmode~\acute{n}\else \'{n}\fi{}ski},\
  and\ \citenamefont {Krajewska}}]{Velez2020}%
  \BibitemOpen
  \bibfield  {author} {\bibinfo {author} {\bibfnamefont {F.}~\bibnamefont
  {Cajiao~V\'elez}}, \bibinfo {author} {\bibfnamefont {J.~Z.}\ \bibnamefont
  {Kami\ifmmode~\acute{n}\else \'{n}\fi{}ski}},\ and\ \bibinfo {author}
  {\bibfnamefont {K.}~\bibnamefont {Krajewska}},\ }\href
  {https://doi.org/10.1103/PhysRevA.101.053430} {\bibfield  {journal} {\bibinfo
   {journal} {Phys. Rev. A}\ }\textbf {\bibinfo {volume} {101}},\ \bibinfo
  {pages} {053430} (\bibinfo {year} {2020})}\BibitemShut {NoStop}%
\bibitem [{\citenamefont {Tolstikhin}\ and\ \citenamefont
  {Morishita}(2019)}]{Tolstikhin2019}%
  \BibitemOpen
  \bibfield  {author} {\bibinfo {author} {\bibfnamefont {O.~I.}\ \bibnamefont
  {Tolstikhin}}\ and\ \bibinfo {author} {\bibfnamefont {T.}~\bibnamefont
  {Morishita}},\ }\href {https://doi.org/10.1103/PhysRevA.99.063415} {\bibfield
   {journal} {\bibinfo  {journal} {Phys. Rev. A}\ }\textbf {\bibinfo {volume}
  {99}},\ \bibinfo {pages} {063415} (\bibinfo {year} {2019})}\BibitemShut
  {NoStop}%
\bibitem [{\citenamefont {Barth}\ and\ \citenamefont
  {Smirnova}(2011)}]{Barth2011}%
  \BibitemOpen
  \bibfield  {author} {\bibinfo {author} {\bibfnamefont {I.}~\bibnamefont
  {Barth}}\ and\ \bibinfo {author} {\bibfnamefont {O.}~\bibnamefont
  {Smirnova}},\ }\href {https://doi.org/10.1103/PhysRevA.84.063415} {\bibfield
  {journal} {\bibinfo  {journal} {Phys. Rev. A}\ }\textbf {\bibinfo {volume}
  {84}},\ \bibinfo {pages} {063415} (\bibinfo {year} {2011})}\BibitemShut
  {NoStop}%
\bibitem [{\citenamefont {Barth}\ and\ \citenamefont
  {Smirnova}(2013)}]{Barth2013}%
  \BibitemOpen
  \bibfield  {author} {\bibinfo {author} {\bibfnamefont {I.}~\bibnamefont
  {Barth}}\ and\ \bibinfo {author} {\bibfnamefont {O.}~\bibnamefont
  {Smirnova}},\ }\href {https://doi.org/10.1103/PhysRevA.87.013433} {\bibfield
  {journal} {\bibinfo  {journal} {Phys. Rev. A}\ }\textbf {\bibinfo {volume}
  {87}},\ \bibinfo {pages} {013433} (\bibinfo {year} {2013})}\BibitemShut
  {NoStop}%
\bibitem [{\citenamefont {{Ngoko Djiokap}}\ \emph {et~al.}(2015)\citenamefont
  {{Ngoko Djiokap}}, \citenamefont {Hu}, \citenamefont {Madsen}, \citenamefont
  {Manakov}, \citenamefont {Meremianin},\ and\ \citenamefont
  {Starace}}]{NgokoDjiokap2015}%
  \BibitemOpen
  \bibfield  {author} {\bibinfo {author} {\bibfnamefont {J.~M.}\ \bibnamefont
  {{Ngoko Djiokap}}}, \bibinfo {author} {\bibfnamefont {S.~X.}\ \bibnamefont
  {Hu}}, \bibinfo {author} {\bibfnamefont {L.~B.}\ \bibnamefont {Madsen}},
  \bibinfo {author} {\bibfnamefont {N.~L.}\ \bibnamefont {Manakov}}, \bibinfo
  {author} {\bibfnamefont {A.~V.}\ \bibnamefont {Meremianin}},\ and\ \bibinfo
  {author} {\bibfnamefont {A.~F.}\ \bibnamefont {Starace}},\ }\href
  {https://doi.org/10.1103/PhysRevLett.115.113004} {\bibfield  {journal}
  {\bibinfo  {journal} {Phys. Rev. Lett.}\ }\textbf {\bibinfo {volume} {115}},\
  \bibinfo {pages} {113004} (\bibinfo {year} {2015})}\BibitemShut {NoStop}%
\bibitem [{\citenamefont {Ngoko~Djiokap}\ \emph {et~al.}(2016)\citenamefont
  {Ngoko~Djiokap}, \citenamefont {Meremianin}, \citenamefont {Manakov},
  \citenamefont {Hu}, \citenamefont {Madsen},\ and\ \citenamefont
  {Starace}}]{NgokoDjiokap2016}%
  \BibitemOpen
  \bibfield  {author} {\bibinfo {author} {\bibfnamefont {J.~M.}\ \bibnamefont
  {Ngoko~Djiokap}}, \bibinfo {author} {\bibfnamefont {A.~V.}\ \bibnamefont
  {Meremianin}}, \bibinfo {author} {\bibfnamefont {N.~L.}\ \bibnamefont
  {Manakov}}, \bibinfo {author} {\bibfnamefont {S.~X.}\ \bibnamefont {Hu}},
  \bibinfo {author} {\bibfnamefont {L.~B.}\ \bibnamefont {Madsen}},\ and\
  \bibinfo {author} {\bibfnamefont {A.~F.}\ \bibnamefont {Starace}},\ }\href
  {https://doi.org/10.1103/PhysRevA.94.013408} {\bibfield  {journal} {\bibinfo
  {journal} {Phys. Rev. A}\ }\textbf {\bibinfo {volume} {94}},\ \bibinfo
  {pages} {013408} (\bibinfo {year} {2016})}\BibitemShut {NoStop}%
\bibitem [{\citenamefont {Yuan}\ \emph {et~al.}(2016)\citenamefont {Yuan},
  \citenamefont {Chelkowski},\ and\ \citenamefont {Bandrauk}}]{yuan2016}%
  \BibitemOpen
  \bibfield  {author} {\bibinfo {author} {\bibfnamefont {K.~J.}\ \bibnamefont
  {Yuan}}, \bibinfo {author} {\bibfnamefont {S.}~\bibnamefont {Chelkowski}},\
  and\ \bibinfo {author} {\bibfnamefont {A.~D.}\ \bibnamefont {Bandrauk}},\
  }\href {https://doi.org/10.1103/PhysRevA.93.053425} {\bibfield  {journal}
  {\bibinfo  {journal} {Phys. Rev. A}\ }\textbf {\bibinfo {volume} {93}},\
  \bibinfo {pages} {053425} (\bibinfo {year} {2016})}\BibitemShut {NoStop}%
\bibitem [{\citenamefont {Ngoko~Djiokap}\ \emph {et~al.}(2018)\citenamefont
  {Ngoko~Djiokap}, \citenamefont {Meremianin}, \citenamefont {Manakov},
  \citenamefont {Madsen}, \citenamefont {Hu},\ and\ \citenamefont
  {Starace}}]{NgokoDjiokap2018}%
  \BibitemOpen
  \bibfield  {author} {\bibinfo {author} {\bibfnamefont {J.~M.}\ \bibnamefont
  {Ngoko~Djiokap}}, \bibinfo {author} {\bibfnamefont {A.~V.}\ \bibnamefont
  {Meremianin}}, \bibinfo {author} {\bibfnamefont {N.~L.}\ \bibnamefont
  {Manakov}}, \bibinfo {author} {\bibfnamefont {L.~B.}\ \bibnamefont {Madsen}},
  \bibinfo {author} {\bibfnamefont {S.~X.}\ \bibnamefont {Hu}},\ and\ \bibinfo
  {author} {\bibfnamefont {A.~F.}\ \bibnamefont {Starace}},\ }\href
  {https://doi.org/10.1103/PhysRevA.98.063407} {\bibfield  {journal} {\bibinfo
  {journal} {Phys. Rev. A}\ }\textbf {\bibinfo {volume} {98}},\ \bibinfo
  {pages} {063407} (\bibinfo {year} {2018})}\BibitemShut {NoStop}%
\bibitem [{\citenamefont {Pengel}\ \emph
  {et~al.}(2017{\natexlab{a}})\citenamefont {Pengel}, \citenamefont
  {Kerbstadt}, \citenamefont {Johannmeyer}, \citenamefont {Englert},
  \citenamefont {Bayer},\ and\ \citenamefont {Wollenhaupt}}]{Pengel2017}%
  \BibitemOpen
  \bibfield  {author} {\bibinfo {author} {\bibfnamefont {D.}~\bibnamefont
  {Pengel}}, \bibinfo {author} {\bibfnamefont {S.}~\bibnamefont {Kerbstadt}},
  \bibinfo {author} {\bibfnamefont {D.}~\bibnamefont {Johannmeyer}}, \bibinfo
  {author} {\bibfnamefont {L.}~\bibnamefont {Englert}}, \bibinfo {author}
  {\bibfnamefont {T.}~\bibnamefont {Bayer}},\ and\ \bibinfo {author}
  {\bibfnamefont {M.}~\bibnamefont {Wollenhaupt}},\ }\href
  {https://doi.org/10.1103/PhysRevLett.118.053003} {\bibfield  {journal}
  {\bibinfo  {journal} {Phys. Rev. Lett.}\ }\textbf {\bibinfo {volume} {118}},\
  \bibinfo {pages} {053003} (\bibinfo {year} {2017}{\natexlab{a}})}\BibitemShut
  {NoStop}%
\bibitem [{\citenamefont {Pengel}\ \emph
  {et~al.}(2017{\natexlab{b}})\citenamefont {Pengel}, \citenamefont
  {Kerbstadt}, \citenamefont {Englert}, \citenamefont {Bayer},\ and\
  \citenamefont {Wollenhaupt}}]{Pengel2017a}%
  \BibitemOpen
  \bibfield  {author} {\bibinfo {author} {\bibfnamefont {D.}~\bibnamefont
  {Pengel}}, \bibinfo {author} {\bibfnamefont {S.}~\bibnamefont {Kerbstadt}},
  \bibinfo {author} {\bibfnamefont {L.}~\bibnamefont {Englert}}, \bibinfo
  {author} {\bibfnamefont {T.}~\bibnamefont {Bayer}},\ and\ \bibinfo {author}
  {\bibfnamefont {M.}~\bibnamefont {Wollenhaupt}},\ }\href
  {https://doi.org/10.1103/PhysRevA.96.043426} {\bibfield  {journal} {\bibinfo
  {journal} {Phys. Rev. A}\ }\textbf {\bibinfo {volume} {96}},\ \bibinfo
  {pages} {043426} (\bibinfo {year} {2017}{\natexlab{b}})}\BibitemShut
  {NoStop}%
\bibitem [{\citenamefont {Li}\ \emph {et~al.}(2018)\citenamefont {Li},
  \citenamefont {Zhang}, \citenamefont {Kong}, \citenamefont {Wang},
  \citenamefont {Ding},\ and\ \citenamefont {Yao}}]{Li2018}%
  \BibitemOpen
  \bibfield  {author} {\bibinfo {author} {\bibfnamefont {M.}~\bibnamefont
  {Li}}, \bibinfo {author} {\bibfnamefont {G.}~\bibnamefont {Zhang}}, \bibinfo
  {author} {\bibfnamefont {X.}~\bibnamefont {Kong}}, \bibinfo {author}
  {\bibfnamefont {T.}~\bibnamefont {Wang}}, \bibinfo {author} {\bibfnamefont
  {X.}~\bibnamefont {Ding}},\ and\ \bibinfo {author} {\bibfnamefont
  {J.}~\bibnamefont {Yao}},\ }\href {https://doi.org/10.1364/oe.26.000878}
  {\bibfield  {journal} {\bibinfo  {journal} {Opt. Express}\ }\textbf {\bibinfo
  {volume} {26}},\ \bibinfo {pages} {878} (\bibinfo {year} {2018})}\BibitemShut
  {NoStop}%
\bibitem [{\citenamefont {Kerbstadt}\ \emph
  {et~al.}(2019{\natexlab{a}})\citenamefont {Kerbstadt}, \citenamefont
  {Eickhoff}, \citenamefont {Bayer},\ and\ \citenamefont
  {Wollenhaupt}}]{Kerbstadt2019}%
  \BibitemOpen
  \bibfield  {author} {\bibinfo {author} {\bibfnamefont {S.}~\bibnamefont
  {Kerbstadt}}, \bibinfo {author} {\bibfnamefont {K.}~\bibnamefont {Eickhoff}},
  \bibinfo {author} {\bibfnamefont {T.}~\bibnamefont {Bayer}},\ and\ \bibinfo
  {author} {\bibfnamefont {M.}~\bibnamefont {Wollenhaupt}},\ }\bibfield
  {journal} {\bibinfo  {journal} {Adv. Phys. X}\ }\textbf {\bibinfo {volume}
  {4}},\ \href {https://doi.org/10.1080/23746149.2019.1672583}
  {10.1080/23746149.2019.1672583} (\bibinfo {year}
  {2019}{\natexlab{a}})\BibitemShut {NoStop}%
\bibitem [{\citenamefont {Kerbstadt}\ \emph
  {et~al.}(2019{\natexlab{b}})\citenamefont {Kerbstadt}, \citenamefont
  {Eickhoff}, \citenamefont {Bayer},\ and\ \citenamefont
  {Wollenhaupt}}]{Kerbstadt2019a}%
  \BibitemOpen
  \bibfield  {author} {\bibinfo {author} {\bibfnamefont {S.}~\bibnamefont
  {Kerbstadt}}, \bibinfo {author} {\bibfnamefont {K.}~\bibnamefont {Eickhoff}},
  \bibinfo {author} {\bibfnamefont {T.}~\bibnamefont {Bayer}},\ and\ \bibinfo
  {author} {\bibfnamefont {M.}~\bibnamefont {Wollenhaupt}},\ }\href
  {https://doi.org/10.1038/s41467-019-08601-7} {\bibfield  {journal} {\bibinfo
  {journal} {Nat. Commun.}\ }\textbf {\bibinfo {volume} {10}},\ \bibinfo
  {pages} {1} (\bibinfo {year} {2019}{\natexlab{b}})}\BibitemShut {NoStop}%
\bibitem [{\citenamefont {Armstrong}\ \emph {et~al.}(2019)\citenamefont
  {Armstrong}, \citenamefont {Clarke}, \citenamefont {Benda}, \citenamefont
  {Wragg}, \citenamefont {Brown},\ and\ \citenamefont {{van der
  Hart}}}]{Armstrong2019b}%
  \BibitemOpen
  \bibfield  {author} {\bibinfo {author} {\bibfnamefont {G.~S.~J.}\
  \bibnamefont {Armstrong}}, \bibinfo {author} {\bibfnamefont {D.~D.~A.}\
  \bibnamefont {Clarke}}, \bibinfo {author} {\bibfnamefont {J.}~\bibnamefont
  {Benda}}, \bibinfo {author} {\bibfnamefont {J.}~\bibnamefont {Wragg}},
  \bibinfo {author} {\bibfnamefont {A.~C.}\ \bibnamefont {Brown}},\ and\
  \bibinfo {author} {\bibfnamefont {H.~W.}\ \bibnamefont {{van der Hart}}},\
  }\href {https://doi.org/10.1103/PhysRevA.100.063416} {\bibfield  {journal}
  {\bibinfo  {journal} {Phys. Rev. A}\ }\textbf {\bibinfo {volume} {100}},\
  \bibinfo {pages} {63416} (\bibinfo {year} {2019})}\BibitemShut {NoStop}%
\bibitem [{\citenamefont {Bayer}\ \emph {et~al.}(2020)\citenamefont {Bayer},
  \citenamefont {Philipp}, \citenamefont {Eickhoff},\ and\ \citenamefont
  {Wollenhaupt}}]{bayer2020}%
  \BibitemOpen
  \bibfield  {author} {\bibinfo {author} {\bibfnamefont {T.}~\bibnamefont
  {Bayer}}, \bibinfo {author} {\bibfnamefont {C.}~\bibnamefont {Philipp}},
  \bibinfo {author} {\bibfnamefont {K.}~\bibnamefont {Eickhoff}},\ and\
  \bibinfo {author} {\bibfnamefont {M.}~\bibnamefont {Wollenhaupt}},\ }\href
  {https://doi.org/10.1103/PhysRevA.102.013104} {\bibfield  {journal} {\bibinfo
   {journal} {Phys. Rev. A}\ }\textbf {\bibinfo {volume} {102}},\ \bibinfo
  {pages} {013104} (\bibinfo {year} {2020})}\BibitemShut {NoStop}%
\bibitem [{\citenamefont {Amini}\ \emph {et~al.}(2019)\citenamefont {Amini},
  \citenamefont {Biegert}, \citenamefont {Calegari}, \citenamefont
  {Chac{\'{o}}n}, \citenamefont {Ciappina}, \citenamefont {Dauphin},
  \citenamefont {Efimov}, \citenamefont {de~Morisson~Faria}, \citenamefont
  {Giergiel}, \citenamefont {Gniewek}, \citenamefont {Landsman}, \citenamefont
  {Lesiuk}, \citenamefont {Mandrysz}, \citenamefont {Maxwell}, \citenamefont
  {Moszy{\'{n}}ski}, \citenamefont {Ortmann}, \citenamefont
  {P{\'{e}}rez-Hern{\'{a}}ndez}, \citenamefont {Pic{\'{o}}n}, \citenamefont
  {Pisanty}, \citenamefont {Prauzner-Bechcicki}, \citenamefont {Sacha},
  \citenamefont {Su{\'{a}}rez}, \citenamefont {Za\"{i}r}, \citenamefont
  {Zakrzewski},\ and\ \citenamefont {Lewenstein}}]{amini2019}%
  \BibitemOpen
  \bibfield  {author} {\bibinfo {author} {\bibfnamefont {K.}~\bibnamefont
  {Amini}}, \bibinfo {author} {\bibfnamefont {J.}~\bibnamefont {Biegert}},
  \bibinfo {author} {\bibfnamefont {F.}~\bibnamefont {Calegari}}, \bibinfo
  {author} {\bibfnamefont {A.}~\bibnamefont {Chac{\'{o}}n}}, \bibinfo {author}
  {\bibfnamefont {M.~F.}\ \bibnamefont {Ciappina}}, \bibinfo {author}
  {\bibfnamefont {A.}~\bibnamefont {Dauphin}}, \bibinfo {author} {\bibfnamefont
  {D.~K.}\ \bibnamefont {Efimov}}, \bibinfo {author} {\bibfnamefont {C.~F.}\
  \bibnamefont {de~Morisson~Faria}}, \bibinfo {author} {\bibfnamefont
  {K.}~\bibnamefont {Giergiel}}, \bibinfo {author} {\bibfnamefont
  {P.}~\bibnamefont {Gniewek}}, \bibinfo {author} {\bibfnamefont {A.~S.}\
  \bibnamefont {Landsman}}, \bibinfo {author} {\bibfnamefont {M.}~\bibnamefont
  {Lesiuk}}, \bibinfo {author} {\bibfnamefont {M.}~\bibnamefont {Mandrysz}},
  \bibinfo {author} {\bibfnamefont {A.~S.}\ \bibnamefont {Maxwell}}, \bibinfo
  {author} {\bibfnamefont {R.}~\bibnamefont {Moszy{\'{n}}ski}}, \bibinfo
  {author} {\bibfnamefont {L.}~\bibnamefont {Ortmann}}, \bibinfo {author}
  {\bibfnamefont {J.~A.}\ \bibnamefont {P{\'{e}}rez-Hern{\'{a}}ndez}}, \bibinfo
  {author} {\bibfnamefont {A.}~\bibnamefont {Pic{\'{o}}n}}, \bibinfo {author}
  {\bibfnamefont {E.}~\bibnamefont {Pisanty}}, \bibinfo {author} {\bibfnamefont
  {J.}~\bibnamefont {Prauzner-Bechcicki}}, \bibinfo {author} {\bibfnamefont
  {K.}~\bibnamefont {Sacha}}, \bibinfo {author} {\bibfnamefont
  {N.}~\bibnamefont {Su{\'{a}}rez}}, \bibinfo {author} {\bibfnamefont
  {A.}~\bibnamefont {Za\"{i}r}}, \bibinfo {author} {\bibfnamefont
  {J.}~\bibnamefont {Zakrzewski}},\ and\ \bibinfo {author} {\bibfnamefont
  {M.}~\bibnamefont {Lewenstein}},\ }\href
  {https://doi.org/10.1088/1361-6633/ab2bb1} {\bibfield  {journal} {\bibinfo
  {journal} {Rep. Prog. Phys.}\ }\textbf {\bibinfo {volume} {82}},\ \bibinfo
  {pages} {116001} (\bibinfo {year} {2019})}\BibitemShut {NoStop}%
\bibitem [{\citenamefont {Mosert}\ and\ \citenamefont {Bauer}(2016)}]{qprop2}%
  \BibitemOpen
  \bibfield  {author} {\bibinfo {author} {\bibfnamefont {V.}~\bibnamefont
  {Mosert}}\ and\ \bibinfo {author} {\bibfnamefont {D.}~\bibnamefont {Bauer}},\
  }\href@noop {} {\bibfield  {journal} {\bibinfo  {journal} {Comp. Phys.
  Commun.}\ }\textbf {\bibinfo {volume} {207}},\ \bibinfo {pages} {452}
  (\bibinfo {year} {2016})}\BibitemShut {NoStop}%
\bibitem [{\citenamefont {Tulsky}\ and\ \citenamefont {Bauer}(2020)}]{qprop3}%
  \BibitemOpen
  \bibfield  {author} {\bibinfo {author} {\bibfnamefont {V.}~\bibnamefont
  {Tulsky}}\ and\ \bibinfo {author} {\bibfnamefont {D.}~\bibnamefont {Bauer}},\
  }\href@noop {} {\bibfield  {journal} {\bibinfo  {journal} {Comp. Phys.
  Commun.}\ }\textbf {\bibinfo {volume} {207}},\ \bibinfo {pages} {107098}
  (\bibinfo {year} {2020})}\BibitemShut {NoStop}%
\bibitem [{\citenamefont {Moore}\ \emph {et~al.}(2011)\citenamefont {Moore},
  \citenamefont {Lysaght}, \citenamefont {Nikolopoulos}, \citenamefont
  {Parker}, \citenamefont {van~der Hart},\ and\ \citenamefont
  {Taylor}}]{moore2011}%
  \BibitemOpen
  \bibfield  {author} {\bibinfo {author} {\bibfnamefont {L.~R.}\ \bibnamefont
  {Moore}}, \bibinfo {author} {\bibfnamefont {M.~A.}\ \bibnamefont {Lysaght}},
  \bibinfo {author} {\bibfnamefont {L.~A.~A.}\ \bibnamefont {Nikolopoulos}},
  \bibinfo {author} {\bibfnamefont {J.~S.}\ \bibnamefont {Parker}}, \bibinfo
  {author} {\bibfnamefont {H.~W.}\ \bibnamefont {van~der Hart}},\ and\ \bibinfo
  {author} {\bibfnamefont {K.~T.}\ \bibnamefont {Taylor}},\ }\href
  {https://doi.org/10.1080/09500340.2011.559315} {\bibfield  {journal}
  {\bibinfo  {journal} {Journal of Modern Optics}\ }\textbf {\bibinfo {volume}
  {58}},\ \bibinfo {pages} {1132} (\bibinfo {year} {2011})}\BibitemShut
  {NoStop}%
\bibitem [{\citenamefont {Clarke}\ \emph {et~al.}(2018)\citenamefont {Clarke},
  \citenamefont {Armstrong}, \citenamefont {Brown},\ and\ \citenamefont
  {van~der Hart}}]{clarke2018}%
  \BibitemOpen
  \bibfield  {author} {\bibinfo {author} {\bibfnamefont {D.~D.~A.}\
  \bibnamefont {Clarke}}, \bibinfo {author} {\bibfnamefont {G.~S.~J.}\
  \bibnamefont {Armstrong}}, \bibinfo {author} {\bibfnamefont {A.~C.}\
  \bibnamefont {Brown}},\ and\ \bibinfo {author} {\bibfnamefont {H.~W.}\
  \bibnamefont {van~der Hart}},\ }\href
  {https://doi.org/10.1103/PhysRevA.98.053442} {\bibfield  {journal} {\bibinfo
  {journal} {Phys. Rev. A}\ }\textbf {\bibinfo {volume} {98}},\ \bibinfo
  {pages} {053442} (\bibinfo {year} {2018})}\BibitemShut {NoStop}%
\bibitem [{\citenamefont {Brown}\ \emph {et~al.}(2020)\citenamefont {Brown},
  \citenamefont {Armstrong}, \citenamefont {Benda}, \citenamefont {Clarke},
  \citenamefont {Wragg}, \citenamefont {Hamilton}, \citenamefont {Mašín},
  \citenamefont {Gorfinkiel},\ and\ \citenamefont {van~der Hart}}]{rmtcpc}%
  \BibitemOpen
  \bibfield  {author} {\bibinfo {author} {\bibfnamefont {A.~C.}\ \bibnamefont
  {Brown}}, \bibinfo {author} {\bibfnamefont {G.~S.~J.}\ \bibnamefont
  {Armstrong}}, \bibinfo {author} {\bibfnamefont {J.}~\bibnamefont {Benda}},
  \bibinfo {author} {\bibfnamefont {D.~D.~A.}\ \bibnamefont {Clarke}}, \bibinfo
  {author} {\bibfnamefont {J.}~\bibnamefont {Wragg}}, \bibinfo {author}
  {\bibfnamefont {K.~R.}\ \bibnamefont {Hamilton}}, \bibinfo {author}
  {\bibfnamefont {Z.}~\bibnamefont {Mašín}}, \bibinfo {author} {\bibfnamefont
  {J.~D.}\ \bibnamefont {Gorfinkiel}},\ and\ \bibinfo {author} {\bibfnamefont
  {H.~W.}\ \bibnamefont {van~der Hart}},\ }\href
  {https://doi.org/https://doi.org/10.1016/j.cpc.2019.107062} {\bibfield
  {journal} {\bibinfo  {journal} {Comput. Phys. Commun.}\ }\textbf {\bibinfo
  {volume} {250}},\ \bibinfo {pages} {107062} (\bibinfo {year}
  {2020})}\BibitemShut {NoStop}%
\bibitem [{\citenamefont {Asenjo-Garcia}\ and\ \citenamefont {Garc\'{\i}a~de
  Abajo}(2014)}]{Asenjo-Garcia2014}%
  \BibitemOpen
  \bibfield  {author} {\bibinfo {author} {\bibfnamefont {A.}~\bibnamefont
  {Asenjo-Garcia}}\ and\ \bibinfo {author} {\bibfnamefont {F.~J.}\ \bibnamefont
  {Garc\'{\i}a~de Abajo}},\ }\href
  {https://doi.org/10.1103/PhysRevLett.113.066102} {\bibfield  {journal}
  {\bibinfo  {journal} {Phys. Rev. Lett.}\ }\textbf {\bibinfo {volume} {113}},\
  \bibinfo {pages} {066102} (\bibinfo {year} {2014})}\BibitemShut {NoStop}%
\bibitem [{\citenamefont {{Becker}}\ \emph {et~al.}(2002)\citenamefont
  {{Becker}}, \citenamefont {{Grasbon}}, \citenamefont {{Kopold}},
  \citenamefont {{Milo{\v s}evi{\'c}}}, \citenamefont {{Paulus}},\ and\
  \citenamefont {{Walther}}}]{Becker2002Review}%
  \BibitemOpen
  \bibfield  {author} {\bibinfo {author} {\bibfnamefont {W.}~\bibnamefont
  {{Becker}}}, \bibinfo {author} {\bibfnamefont {F.}~\bibnamefont {{Grasbon}}},
  \bibinfo {author} {\bibfnamefont {R.}~\bibnamefont {{Kopold}}}, \bibinfo
  {author} {\bibfnamefont {D.~B.}\ \bibnamefont {{Milo{\v s}evi{\'c}}}},
  \bibinfo {author} {\bibfnamefont {G.~G.}\ \bibnamefont {{Paulus}}},\ and\
  \bibinfo {author} {\bibfnamefont {H.}~\bibnamefont {{Walther}}},\ }\href
  {https://doi.org/10.1016/S1049-250X(02)80006-4} {\bibfield  {journal}
  {\bibinfo  {journal} {Adv. At. Mol. Opt. Phys.}\ }\textbf {\bibinfo {volume}
  {48}},\ \bibinfo {pages} {35} (\bibinfo {year} {2002})}\BibitemShut {NoStop}%
\bibitem [{\citenamefont {Figueira~de Morisson~Faria}\ \emph
  {et~al.}(2002)\citenamefont {Figueira~de Morisson~Faria}, \citenamefont
  {Schomerus},\ and\ \citenamefont {Becker}}]{Faria2002}%
  \BibitemOpen
  \bibfield  {author} {\bibinfo {author} {\bibfnamefont {C.}~\bibnamefont
  {Figueira~de Morisson~Faria}}, \bibinfo {author} {\bibfnamefont
  {H.}~\bibnamefont {Schomerus}},\ and\ \bibinfo {author} {\bibfnamefont
  {W.}~\bibnamefont {Becker}},\ }\href
  {https://doi.org/10.1103/PhysRevA.66.043413} {\bibfield  {journal} {\bibinfo
  {journal} {Phys. Rev. A}\ }\textbf {\bibinfo {volume} {66}},\ \bibinfo
  {pages} {043413} (\bibinfo {year} {2002})}\BibitemShut {NoStop}%
\bibitem [{\citenamefont {Keldysh}(1965)}]{Keldysh1965}%
  \BibitemOpen
  \bibfield  {author} {\bibinfo {author} {\bibfnamefont {L.~V.}\ \bibnamefont
  {Keldysh}},\ }\href {https://doi.org/10.1234/12345678} {\bibfield  {journal}
  {\bibinfo  {journal} {Sov. Phys. JETP}\ }\textbf {\bibinfo {volume} {20}},\
  \bibinfo {pages} {1307} (\bibinfo {year} {1965})}\BibitemShut {NoStop}%
\bibitem [{\citenamefont {Faisal}(1973)}]{Faisal1973}%
  \BibitemOpen
  \bibfield  {author} {\bibinfo {author} {\bibfnamefont {F.~H.}\ \bibnamefont
  {Faisal}},\ }\href {https://doi.org/10.1088/0022-3700/6/4/011} {\bibfield
  {journal} {\bibinfo  {journal} {J. Phys. B: At. Mol. Phys.}\ }\textbf
  {\bibinfo {volume} {6}},\ \bibinfo {pages} {L89} (\bibinfo {year}
  {1973})}\BibitemShut {NoStop}%
\bibitem [{\citenamefont {Reiss}(1980)}]{Reiss1980}%
  \BibitemOpen
  \bibfield  {author} {\bibinfo {author} {\bibfnamefont {H.~R.}\ \bibnamefont
  {Reiss}},\ }\href {https://doi.org/10.1103/PhysRevA.22.1786} {\bibfield
  {journal} {\bibinfo  {journal} {Phys. Rev. A}\ }\textbf {\bibinfo {volume}
  {22}},\ \bibinfo {pages} {1786} (\bibinfo {year} {1980})}\BibitemShut
  {NoStop}%
\bibitem [{\citenamefont {Tong}\ and\ \citenamefont {Lin}(2005)}]{tong}%
  \BibitemOpen
  \bibfield  {author} {\bibinfo {author} {\bibfnamefont {X.~M.}\ \bibnamefont
  {Tong}}\ and\ \bibinfo {author} {\bibfnamefont {C.~D.}\ \bibnamefont {Lin}},\
  }\href@noop {} {\bibfield  {journal} {\bibinfo  {journal} {J. Phys. B: At.
  Mol. Opt. Phys.}\ }\textbf {\bibinfo {volume} {38}},\ \bibinfo {pages} {2593}
  (\bibinfo {year} {2005})}\BibitemShut {NoStop}%
\bibitem [{\citenamefont {Burke}\ and\ \citenamefont
  {Taylor}(1975)}]{burke1975}%
  \BibitemOpen
  \bibfield  {author} {\bibinfo {author} {\bibfnamefont {P.~G.}\ \bibnamefont
  {Burke}}\ and\ \bibinfo {author} {\bibfnamefont {K.~T.}\ \bibnamefont
  {Taylor}},\ }\href {https://doi.org/10.1088/0022-3700/8/16/020} {\bibfield
  {journal} {\bibinfo  {journal} {J. Phys. B: At. Mol. Opt. Phys.}\ }\textbf
  {\bibinfo {volume} {8}},\ \bibinfo {pages} {2620} (\bibinfo {year}
  {1975})}\BibitemShut {NoStop}%
\bibitem [{\citenamefont {Lewenstein}\ \emph {et~al.}(1995)\citenamefont
  {Lewenstein}, \citenamefont {Kulander}, \citenamefont {Schafer},\ and\
  \citenamefont {Bucksbaum}}]{Lewenstein1995}%
  \BibitemOpen
  \bibfield  {author} {\bibinfo {author} {\bibfnamefont {M.}~\bibnamefont
  {Lewenstein}}, \bibinfo {author} {\bibfnamefont {K.~C.}\ \bibnamefont
  {Kulander}}, \bibinfo {author} {\bibfnamefont {K.~J.}\ \bibnamefont
  {Schafer}},\ and\ \bibinfo {author} {\bibfnamefont {P.~H.}\ \bibnamefont
  {Bucksbaum}},\ }\href {https://doi.org/10.1103/PhysRevA.51.1495} {\bibfield
  {journal} {\bibinfo  {journal} {Phys. Rev. A}\ }\textbf {\bibinfo {volume}
  {51}},\ \bibinfo {pages} {1495} (\bibinfo {year} {1995})}\BibitemShut
  {NoStop}%
\bibitem [{\citenamefont {Fetter}(2009)}]{Fetter09}%
  \BibitemOpen
  \bibfield  {author} {\bibinfo {author} {\bibfnamefont {A.}~\bibnamefont
  {Fetter}},\ }\href@noop {} {\bibfield  {journal} {\bibinfo  {journal} {Rev.
  Mod. Phys.}\ }\textbf {\bibinfo {volume} {81}},\ \bibinfo {pages} {647}
  (\bibinfo {year} {2009})}\BibitemShut {NoStop}%
\bibitem [{\citenamefont {Pitaevskii}\ and\ \citenamefont
  {Stringari}(2016)}]{PitaevskiiStringari2016book}%
  \BibitemOpen
  \bibfield  {author} {\bibinfo {author} {\bibfnamefont {L.}~\bibnamefont
  {Pitaevskii}}\ and\ \bibinfo {author} {\bibfnamefont {S.}~\bibnamefont
  {Stringari}},\ }\href
  {https://global.oup.com/academic/product/bose-einstein-condensation-and-superfluidity-9780198758884}
  {\emph {\bibinfo {title} {Bose-Einstein Condensation and Superfluidity}}},\
  \bibinfo {edition} {2nd}\ ed.\ (\bibinfo  {publisher} {Oxford Univ. Press},\
  \bibinfo {address} {Oxford},\ \bibinfo {year} {2016})\BibitemShut {NoStop}%
\bibitem [{\citenamefont {Madison}\ \emph {et~al.}(2000)\citenamefont
  {Madison}, \citenamefont {Chevy}, \citenamefont {Wohlleben},\ and\
  \citenamefont {Dalibard}}]{Madison00}%
  \BibitemOpen
  \bibfield  {author} {\bibinfo {author} {\bibfnamefont {K.~W.}\ \bibnamefont
  {Madison}}, \bibinfo {author} {\bibfnamefont {F.}~\bibnamefont {Chevy}},
  \bibinfo {author} {\bibfnamefont {W.}~\bibnamefont {Wohlleben}},\ and\
  \bibinfo {author} {\bibfnamefont {J.}~\bibnamefont {Dalibard}},\ }\href
  {https://doi.org/10.1103/PhysRevLett.84.806} {\bibfield  {journal} {\bibinfo
  {journal} {Phys. Rev. Lett.}\ }\textbf {\bibinfo {volume} {84}},\ \bibinfo
  {pages} {806} (\bibinfo {year} {2000})}\BibitemShut {NoStop}%
\bibitem [{\citenamefont {Srinivasan}(2006)}]{Srinivasan06}%
  \BibitemOpen
  \bibfield  {author} {\bibinfo {author} {\bibfnamefont {R.}~\bibnamefont
  {Srinivasan}},\ }\href {https://doi.org/10.1007/BF02704934} {\bibfield
  {journal} {\bibinfo  {journal} {Pramana}\ }\textbf {\bibinfo {volume} {66}},\
  \bibinfo {pages} {3} (\bibinfo {year} {2006})}\BibitemShut {NoStop}%
\bibitem [{\citenamefont {Dobrek}\ \emph {et~al.}(1999)\citenamefont {Dobrek},
  \citenamefont {Gajda}, \citenamefont {Lewenstein}, \citenamefont {Sengstock},
  \citenamefont {Birkl},\ and\ \citenamefont {Ertmer}}]{Dobrek1999}%
  \BibitemOpen
  \bibfield  {author} {\bibinfo {author} {\bibfnamefont {{\L{}}.}~\bibnamefont
  {Dobrek}}, \bibinfo {author} {\bibfnamefont {M.}~\bibnamefont {Gajda}},
  \bibinfo {author} {\bibfnamefont {M.}~\bibnamefont {Lewenstein}}, \bibinfo
  {author} {\bibfnamefont {K.}~\bibnamefont {Sengstock}}, \bibinfo {author}
  {\bibfnamefont {G.}~\bibnamefont {Birkl}},\ and\ \bibinfo {author}
  {\bibfnamefont {W.}~\bibnamefont {Ertmer}},\ }\href
  {https://doi.org/10.1103/PhysRevA.60.R3381} {\bibfield  {journal} {\bibinfo
  {journal} {Phys. Rev. A}\ }\textbf {\bibinfo {volume} {60}},\ \bibinfo
  {pages} {R3381} (\bibinfo {year} {1999})}\BibitemShut {NoStop}%
\bibitem [{\citenamefont {Bolda}\ and\ \citenamefont {Walls}(1998)}]{Bolda98}%
  \BibitemOpen
  \bibfield  {author} {\bibinfo {author} {\bibfnamefont {E.~L.}\ \bibnamefont
  {Bolda}}\ and\ \bibinfo {author} {\bibfnamefont {D.~F.}\ \bibnamefont
  {Walls}},\ }\href {https://doi.org/10.1103/PhysRevLett.81.5477} {\bibfield
  {journal} {\bibinfo  {journal} {Phys. Rev. Lett.}\ }\textbf {\bibinfo
  {volume} {81}},\ \bibinfo {pages} {5477} (\bibinfo {year}
  {1998})}\BibitemShut {NoStop}%
\bibitem [{\citenamefont {Tempere}\ and\ \citenamefont
  {Devreese}(1998)}]{Tempere98}%
  \BibitemOpen
  \bibfield  {author} {\bibinfo {author} {\bibfnamefont {J.}~\bibnamefont
  {Tempere}}\ and\ \bibinfo {author} {\bibfnamefont {J.~T.}\ \bibnamefont
  {Devreese}},\ }\href@noop {} {\bibfield  {journal} {\bibinfo  {journal}
  {Solid State Commun.}\ }\textbf {\bibinfo {volume} {108}},\ \bibinfo {pages}
  {993} (\bibinfo {year} {1998})}\BibitemShut {NoStop}%
\bibitem [{\citenamefont {Dalfovo}\ \emph {et~al.}(1999)\citenamefont
  {Dalfovo}, \citenamefont {Giorgini}, \citenamefont {Pitaevskii},\ and\
  \citenamefont {Stringari}}]{Dalfovo99}%
  \BibitemOpen
  \bibfield  {author} {\bibinfo {author} {\bibfnamefont {F.}~\bibnamefont
  {Dalfovo}}, \bibinfo {author} {\bibfnamefont {S.}~\bibnamefont {Giorgini}},
  \bibinfo {author} {\bibfnamefont {L.~P.}\ \bibnamefont {Pitaevskii}},\ and\
  \bibinfo {author} {\bibfnamefont {S.}~\bibnamefont {Stringari}},\ }\href
  {https://doi.org/10.1103/RevModPhys.71.463} {\bibfield  {journal} {\bibinfo
  {journal} {Rev. Mod. Phys.}\ }\textbf {\bibinfo {volume} {71}},\ \bibinfo
  {pages} {463} (\bibinfo {year} {1999})}\BibitemShut {NoStop}%
\bibitem [{\citenamefont {Staliunas}\ \emph {et~al.}(1999)\citenamefont
  {Staliunas}, \citenamefont {Weiss},\ and\ \citenamefont
  {Slekys}}]{Staliunas99}%
  \BibitemOpen
  \bibfield  {author} {\bibinfo {author} {\bibfnamefont {K.}~\bibnamefont
  {Staliunas}}, \bibinfo {author} {\bibfnamefont {C.~O.}\ \bibnamefont
  {Weiss}},\ and\ \bibinfo {author} {\bibfnamefont {G.}~\bibnamefont
  {Slekys}},\ }\href@noop {} {\emph {\bibinfo {title} {Horizons of World
  Physics}}},\ Vol.\ \bibinfo {volume} {228}\ (\bibinfo  {publisher} {Nova
  Science Publishers, Commack, NY},\ \bibinfo {year} {1999})\BibitemShut
  {NoStop}%
\bibitem [{\citenamefont {Basistiy}\ \emph {et~al.}(1993)\citenamefont
  {Basistiy}, \citenamefont {Bazhenov}, \citenamefont {Soskin},\ and\
  \citenamefont {Vasnetsov}}]{Basistiy93}%
  \BibitemOpen
  \bibfield  {author} {\bibinfo {author} {\bibfnamefont {I.}~\bibnamefont
  {Basistiy}}, \bibinfo {author} {\bibfnamefont {V.}~\bibnamefont {Bazhenov}},
  \bibinfo {author} {\bibfnamefont {M.}~\bibnamefont {Soskin}},\ and\ \bibinfo
  {author} {\bibfnamefont {M.}~\bibnamefont {Vasnetsov}},\ }\href
  {https://doi.org/https://doi.org/10.1016/0030-4018(93)90168-5} {\bibfield
  {journal} {\bibinfo  {journal} {Opt. Commun.}\ }\textbf {\bibinfo {volume}
  {103}},\ \bibinfo {pages} {422 } (\bibinfo {year} {1993})}\BibitemShut
  {NoStop}%
\bibitem [{\citenamefont {Kreminskaya}\ \emph {et~al.}(1998)\citenamefont
  {Kreminskaya}, \citenamefont {Soskin},\ and\ \citenamefont
  {Khizhnyak}}]{Kreminskaya98}%
  \BibitemOpen
  \bibfield  {author} {\bibinfo {author} {\bibfnamefont {L.~V.}\ \bibnamefont
  {Kreminskaya}}, \bibinfo {author} {\bibfnamefont {M.~S.}\ \bibnamefont
  {Soskin}},\ and\ \bibinfo {author} {\bibfnamefont {A.~I.}\ \bibnamefont
  {Khizhnyak}},\ }\href@noop {} {\bibfield  {journal} {\bibinfo  {journal}
  {Opt. Commun.}\ }\textbf {\bibinfo {volume} {145}},\ \bibinfo {pages} {377}
  (\bibinfo {year} {1998})}\BibitemShut {NoStop}%
\bibitem [{\citenamefont {Nye}\ and\ \citenamefont {Berry}(1974)}]{Nye74}%
  \BibitemOpen
  \bibfield  {author} {\bibinfo {author} {\bibfnamefont {J.~F.}\ \bibnamefont
  {Nye}}\ and\ \bibinfo {author} {\bibfnamefont {M.~V.}\ \bibnamefont
  {Berry}},\ }\href@noop {} {\bibfield  {journal} {\bibinfo  {journal} {Proc.
  R. Soc. London, Ser. A}\ }\textbf {\bibinfo {volume} {336}},\ \bibinfo
  {pages} {165} (\bibinfo {year} {1974})}\BibitemShut {NoStop}%
\bibitem [{\citenamefont {Boutu}\ \emph {et~al.}(2011)\citenamefont {Boutu},
  \citenamefont {Auguste}, \citenamefont {Boyko}, \citenamefont {Sola},
  \citenamefont {Balcou}, \citenamefont {Binazon}, \citenamefont {Gobert},
  \citenamefont {Merdji}, \citenamefont {Valentin}, \citenamefont {Constant},
  \citenamefont {M\'evel},\ and\ \citenamefont {Carr\'e}}]{Boutu2011}%
  \BibitemOpen
  \bibfield  {author} {\bibinfo {author} {\bibfnamefont {W.}~\bibnamefont
  {Boutu}}, \bibinfo {author} {\bibfnamefont {T.}~\bibnamefont {Auguste}},
  \bibinfo {author} {\bibfnamefont {O.}~\bibnamefont {Boyko}}, \bibinfo
  {author} {\bibfnamefont {I.}~\bibnamefont {Sola}}, \bibinfo {author}
  {\bibfnamefont {P.}~\bibnamefont {Balcou}}, \bibinfo {author} {\bibfnamefont
  {L.}~\bibnamefont {Binazon}}, \bibinfo {author} {\bibfnamefont
  {O.}~\bibnamefont {Gobert}}, \bibinfo {author} {\bibfnamefont
  {H.}~\bibnamefont {Merdji}}, \bibinfo {author} {\bibfnamefont
  {C.}~\bibnamefont {Valentin}}, \bibinfo {author} {\bibfnamefont
  {E.}~\bibnamefont {Constant}}, \bibinfo {author} {\bibfnamefont
  {E.}~\bibnamefont {M\'evel}},\ and\ \bibinfo {author} {\bibfnamefont
  {B.}~\bibnamefont {Carr\'e}},\ }\href
  {https://doi.org/10.1103/PhysRevA.84.063406} {\bibfield  {journal} {\bibinfo
  {journal} {Phys. Rev. A}\ }\textbf {\bibinfo {volume} {84}},\ \bibinfo
  {pages} {063406} (\bibinfo {year} {2011})}\BibitemShut {NoStop}%
\bibitem [{\citenamefont {Toma}\ \emph {et~al.}(1999)\citenamefont {Toma},
  \citenamefont {Antoine}, \citenamefont {de~Bohan},\ and\ \citenamefont
  {Muller}}]{Toma1999}%
  \BibitemOpen
  \bibfield  {author} {\bibinfo {author} {\bibfnamefont {E.~S.}\ \bibnamefont
  {Toma}}, \bibinfo {author} {\bibfnamefont {P.}~\bibnamefont {Antoine}},
  \bibinfo {author} {\bibfnamefont {A.}~\bibnamefont {de~Bohan}},\ and\
  \bibinfo {author} {\bibfnamefont {H.~G.}\ \bibnamefont {Muller}},\ }\href
  {https://doi.org/10.1088/0953-4075/32/24/318} {\bibfield  {journal} {\bibinfo
   {journal} {J. Phys. B: At. Mol. Opt. Phys.}\ }\textbf {\bibinfo {volume}
  {32}},\ \bibinfo {pages} {5843} (\bibinfo {year} {1999})}\BibitemShut
  {NoStop}%
\bibitem [{\citenamefont {Eppink}\ and\ \citenamefont
  {Parker}(1997)}]{Eppink1997}%
  \BibitemOpen
  \bibfield  {author} {\bibinfo {author} {\bibfnamefont {A.~T. J.~B.}\
  \bibnamefont {Eppink}}\ and\ \bibinfo {author} {\bibfnamefont {D.~H.}\
  \bibnamefont {Parker}},\ }\href {https://doi.org/10.1063/1.1148310}
  {\bibfield  {journal} {\bibinfo  {journal} {Rev. Sci. Instrum.}\ }\textbf
  {\bibinfo {volume} {68}},\ \bibinfo {pages} {3477} (\bibinfo {year}
  {1997})},\ \Eprint {https://arxiv.org/abs/https://doi.org/10.1063/1.1148310}
  {https://doi.org/10.1063/1.1148310} \BibitemShut {NoStop}%
\bibitem [{\citenamefont {Takahashi}\ \emph {et~al.}(2000)\citenamefont
  {Takahashi}, \citenamefont {Cave},\ and\ \citenamefont
  {Eland}}]{Takahashi2000}%
  \BibitemOpen
  \bibfield  {author} {\bibinfo {author} {\bibfnamefont {M.}~\bibnamefont
  {Takahashi}}, \bibinfo {author} {\bibfnamefont {J.~P.}\ \bibnamefont
  {Cave}},\ and\ \bibinfo {author} {\bibfnamefont {J.~H.~D.}\ \bibnamefont
  {Eland}},\ }\href {https://doi.org/10.1063/1.1150460} {\bibfield  {journal}
  {\bibinfo  {journal} {Rev. Sci. Instrum.}\ }\textbf {\bibinfo {volume}
  {71}},\ \bibinfo {pages} {1337} (\bibinfo {year} {2000})},\ \Eprint
  {https://arxiv.org/abs/https://doi.org/10.1063/1.1150460}
  {https://doi.org/10.1063/1.1150460} \BibitemShut {NoStop}%
\bibitem [{\citenamefont {Moshammer}\ \emph {et~al.}(1996)\citenamefont
  {Moshammer}, \citenamefont {Unverzagt}, \citenamefont {Schmitt},
  \citenamefont {Ullrich},\ and\ \citenamefont
  {Schmidt-Böcking}}]{Moshammer1996}%
  \BibitemOpen
  \bibfield  {author} {\bibinfo {author} {\bibfnamefont {R.}~\bibnamefont
  {Moshammer}}, \bibinfo {author} {\bibfnamefont {M.}~\bibnamefont
  {Unverzagt}}, \bibinfo {author} {\bibfnamefont {W.}~\bibnamefont {Schmitt}},
  \bibinfo {author} {\bibfnamefont {J.}~\bibnamefont {Ullrich}},\ and\ \bibinfo
  {author} {\bibfnamefont {H.}~\bibnamefont {Schmidt-Böcking}},\ }\href
  {https://doi.org/https://doi.org/10.1016/0168-583X(95)01259-1} {\bibfield
  {journal} {\bibinfo  {journal} {Nucl. Instrum. Methods Phys. Res. B}\
  }\textbf {\bibinfo {volume} {108}},\ \bibinfo {pages} {425 } (\bibinfo {year}
  {1996})}\BibitemShut {NoStop}%
\bibitem [{\citenamefont {Dörner}\ \emph {et~al.}(2000)\citenamefont
  {Dörner}, \citenamefont {Mergel}, \citenamefont {Jagutzki}, \citenamefont
  {Spielberger}, \citenamefont {Ullrich}, \citenamefont {Moshammer},\ and\
  \citenamefont {Schmidt-Böcking}}]{Dorner2000}%
  \BibitemOpen
  \bibfield  {author} {\bibinfo {author} {\bibfnamefont {R.}~\bibnamefont
  {Dörner}}, \bibinfo {author} {\bibfnamefont {V.}~\bibnamefont {Mergel}},
  \bibinfo {author} {\bibfnamefont {O.}~\bibnamefont {Jagutzki}}, \bibinfo
  {author} {\bibfnamefont {L.}~\bibnamefont {Spielberger}}, \bibinfo {author}
  {\bibfnamefont {J.}~\bibnamefont {Ullrich}}, \bibinfo {author} {\bibfnamefont
  {R.}~\bibnamefont {Moshammer}},\ and\ \bibinfo {author} {\bibfnamefont
  {H.}~\bibnamefont {Schmidt-Böcking}},\ }\href
  {https://doi.org/https://doi.org/10.1016/S0370-1573(99)00109-X} {\bibfield
  {journal} {\bibinfo  {journal} {Phys. Rep.}\ }\textbf {\bibinfo {volume}
  {330}},\ \bibinfo {pages} {95 } (\bibinfo {year} {2000})}\BibitemShut
  {NoStop}%
\bibitem [{\citenamefont {Ullrich}\ \emph {et~al.}(2003)\citenamefont
  {Ullrich}, \citenamefont {Moshammer}, \citenamefont {Dorn}, \citenamefont
  {rner}, \citenamefont {Schmidt},\ and\ \citenamefont {cking}}]{Ullrich2003}%
  \BibitemOpen
  \bibfield  {author} {\bibinfo {author} {\bibfnamefont {J.}~\bibnamefont
  {Ullrich}}, \bibinfo {author} {\bibfnamefont {R.}~\bibnamefont {Moshammer}},
  \bibinfo {author} {\bibfnamefont {A.}~\bibnamefont {Dorn}}, \bibinfo {author}
  {\bibfnamefont {R.~D.}\ \bibnamefont {rner}}, \bibinfo {author}
  {\bibfnamefont {L.~P.~H.}\ \bibnamefont {Schmidt}},\ and\ \bibinfo {author}
  {\bibfnamefont {H.~S.-B.}\ \bibnamefont {cking}},\ }\href
  {https://doi.org/10.1088/0034-4885/66/9/203} {\bibfield  {journal} {\bibinfo
  {journal} {Rep. Prog. Phys.}\ }\textbf {\bibinfo {volume} {66}},\ \bibinfo
  {pages} {1463} (\bibinfo {year} {2003})}\BibitemShut {NoStop}%
\bibitem [{\citenamefont {Heinosaari}\ \emph {et~al.}(2016)\citenamefont
  {Heinosaari}, \citenamefont {Miyadera},\ and\ \citenamefont
  {Ziman}}]{Heinosaari2016}%
  \BibitemOpen
  \bibfield  {author} {\bibinfo {author} {\bibfnamefont {T.}~\bibnamefont
  {Heinosaari}}, \bibinfo {author} {\bibfnamefont {T.}~\bibnamefont
  {Miyadera}},\ and\ \bibinfo {author} {\bibfnamefont {M.}~\bibnamefont
  {Ziman}},\ }\href {https://doi.org/10.1088/1751-8113/49/12/123001} {\bibfield
   {journal} {\bibinfo  {journal} {J. Phys. A: Math. Theor.}\ }\textbf
  {\bibinfo {volume} {49}},\ \bibinfo {pages} {123001} (\bibinfo {year}
  {2016})}\BibitemShut {NoStop}%
\bibitem [{\citenamefont {Bongs}\ \emph {et~al.}(2019)\citenamefont {Bongs},
  \citenamefont {Holynski}, \citenamefont {Vovrosh}, \citenamefont {Bouyer},
  \citenamefont {Condon}, \citenamefont {Rasel}, \citenamefont {Schubert},
  \citenamefont {Schleich},\ and\ \citenamefont {Roura}}]{Bongs2019}%
  \BibitemOpen
  \bibfield  {author} {\bibinfo {author} {\bibfnamefont {K.}~\bibnamefont
  {Bongs}}, \bibinfo {author} {\bibfnamefont {M.}~\bibnamefont {Holynski}},
  \bibinfo {author} {\bibfnamefont {J.}~\bibnamefont {Vovrosh}}, \bibinfo
  {author} {\bibfnamefont {P.}~\bibnamefont {Bouyer}}, \bibinfo {author}
  {\bibfnamefont {G.}~\bibnamefont {Condon}}, \bibinfo {author} {\bibfnamefont
  {E.}~\bibnamefont {Rasel}}, \bibinfo {author} {\bibfnamefont
  {C.}~\bibnamefont {Schubert}}, \bibinfo {author} {\bibfnamefont {W.~P.}\
  \bibnamefont {Schleich}},\ and\ \bibinfo {author} {\bibfnamefont
  {A.}~\bibnamefont {Roura}},\ }\href
  {https://doi.org/10.1038/s42254-019-0117-4} {\bibfield  {journal} {\bibinfo
  {journal} {Nat. Rev. Phys.}\ }\textbf {\bibinfo {volume} {1}},\ \bibinfo
  {pages} {731} (\bibinfo {year} {2019})}\BibitemShut {NoStop}%
\bibitem [{\citenamefont {Faria}\ and\ \citenamefont
  {Maxwell}(2020)}]{faria_it_2020}%
  \BibitemOpen
  \bibfield  {author} {\bibinfo {author} {\bibfnamefont {C.~F. d.~M.}\
  \bibnamefont {Faria}}\ and\ \bibinfo {author} {\bibfnamefont {A.~S.}\
  \bibnamefont {Maxwell}},\ }\href {https://doi.org/10.1088/1361-6633/ab5c91}
  {\bibfield  {journal} {\bibinfo  {journal} {Rep Prog. Phys.}\ }\textbf
  {\bibinfo {volume} {83}},\ \bibinfo {pages} {034401} (\bibinfo {year}
  {2020})},\ \bibinfo {note} {publisher: IOP Publishing}\BibitemShut {NoStop}%
\bibitem [{\citenamefont {Guzzinati}\ \emph {et~al.}(2014)\citenamefont
  {Guzzinati}, \citenamefont {Clark}, \citenamefont {B\'ech\'e},\ and\
  \citenamefont {Verbeeck}}]{Guzzinati2014}%
  \BibitemOpen
  \bibfield  {author} {\bibinfo {author} {\bibfnamefont {G.}~\bibnamefont
  {Guzzinati}}, \bibinfo {author} {\bibfnamefont {L.}~\bibnamefont {Clark}},
  \bibinfo {author} {\bibfnamefont {A.}~\bibnamefont {B\'ech\'e}},\ and\
  \bibinfo {author} {\bibfnamefont {J.}~\bibnamefont {Verbeeck}},\ }\href
  {https://doi.org/10.1103/PhysRevA.89.025803} {\bibfield  {journal} {\bibinfo
  {journal} {Phys. Rev. A}\ }\textbf {\bibinfo {volume} {89}},\ \bibinfo
  {pages} {025803} (\bibinfo {year} {2014})}\BibitemShut {NoStop}%
\bibitem [{\citenamefont {Saitoh}\ \emph {et~al.}(2013)\citenamefont {Saitoh},
  \citenamefont {Hasegawa}, \citenamefont {Hirakawa}, \citenamefont {Tanaka},\
  and\ \citenamefont {Uchida}}]{Saitoh2013}%
  \BibitemOpen
  \bibfield  {author} {\bibinfo {author} {\bibfnamefont {K.}~\bibnamefont
  {Saitoh}}, \bibinfo {author} {\bibfnamefont {Y.}~\bibnamefont {Hasegawa}},
  \bibinfo {author} {\bibfnamefont {K.}~\bibnamefont {Hirakawa}}, \bibinfo
  {author} {\bibfnamefont {N.}~\bibnamefont {Tanaka}},\ and\ \bibinfo {author}
  {\bibfnamefont {M.}~\bibnamefont {Uchida}},\ }\href
  {https://doi.org/10.1103/PhysRevLett.111.074801} {\bibfield  {journal}
  {\bibinfo  {journal} {Phys. Rev. Lett.}\ }\textbf {\bibinfo {volume} {111}},\
  \bibinfo {pages} {074801} (\bibinfo {year} {2013})}\BibitemShut {NoStop}%
\bibitem [{\citenamefont {Grillo}\ \emph {et~al.}(2017)\citenamefont {Grillo},
  \citenamefont {Tavabi}, \citenamefont {Venturi}, \citenamefont {Larocque},
  \citenamefont {Balboni}, \citenamefont {Gazzadi}, \citenamefont {Frabboni},
  \citenamefont {Lu}, \citenamefont {Mafakheri}, \citenamefont {Bouchard},
  \citenamefont {Dunin-Borkowski}, \citenamefont {Boyd}, \citenamefont
  {Lavery}, \citenamefont {Padgett},\ and\ \citenamefont
  {Karimi}}]{Grillo2017}%
  \BibitemOpen
  \bibfield  {author} {\bibinfo {author} {\bibfnamefont {V.}~\bibnamefont
  {Grillo}}, \bibinfo {author} {\bibfnamefont {A.~H.}\ \bibnamefont {Tavabi}},
  \bibinfo {author} {\bibfnamefont {F.}~\bibnamefont {Venturi}}, \bibinfo
  {author} {\bibfnamefont {H.}~\bibnamefont {Larocque}}, \bibinfo {author}
  {\bibfnamefont {R.}~\bibnamefont {Balboni}}, \bibinfo {author} {\bibfnamefont
  {G.~C.}\ \bibnamefont {Gazzadi}}, \bibinfo {author} {\bibfnamefont
  {S.}~\bibnamefont {Frabboni}}, \bibinfo {author} {\bibfnamefont {P.~H.}\
  \bibnamefont {Lu}}, \bibinfo {author} {\bibfnamefont {E.}~\bibnamefont
  {Mafakheri}}, \bibinfo {author} {\bibfnamefont {F.}~\bibnamefont {Bouchard}},
  \bibinfo {author} {\bibfnamefont {R.~E.}\ \bibnamefont {Dunin-Borkowski}},
  \bibinfo {author} {\bibfnamefont {R.~W.}\ \bibnamefont {Boyd}}, \bibinfo
  {author} {\bibfnamefont {M.~P.~J.}\ \bibnamefont {Lavery}}, \bibinfo {author}
  {\bibfnamefont {M.~J.}\ \bibnamefont {Padgett}},\ and\ \bibinfo {author}
  {\bibfnamefont {E.}~\bibnamefont {Karimi}},\ }\href
  {https://doi.org/10.1038/ncomms15536} {\bibfield  {journal} {\bibinfo
  {journal} {Nat Commun.}\ }\textbf {\bibinfo {volume} {8}},\ \bibinfo {pages}
  {15536} (\bibinfo {year} {2017})}\BibitemShut {NoStop}%
\bibitem [{\citenamefont {Dorney}\ \emph {et~al.}(2019)\citenamefont {Dorney},
  \citenamefont {Rego}, \citenamefont {Brooks}, \citenamefont {San~Rom{\'a}n},
  \citenamefont {Liao}, \citenamefont {Ellis}, \citenamefont {Zusin},
  \citenamefont {Gentry}, \citenamefont {Nguyen}, \citenamefont {Shaw},
  \citenamefont {Pic{\'o}n}, \citenamefont {Plaja}, \citenamefont {Kapteyn},
  \citenamefont {Murnane},\ and\ \citenamefont
  {Hern{\'a}ndez-Garc{\'i}a}}]{Dorney2019}%
  \BibitemOpen
  \bibfield  {author} {\bibinfo {author} {\bibfnamefont {K.~M.}\ \bibnamefont
  {Dorney}}, \bibinfo {author} {\bibfnamefont {L.}~\bibnamefont {Rego}},
  \bibinfo {author} {\bibfnamefont {N.~J.}\ \bibnamefont {Brooks}}, \bibinfo
  {author} {\bibfnamefont {J.}~\bibnamefont {San~Rom{\'a}n}}, \bibinfo {author}
  {\bibfnamefont {C.-T.}\ \bibnamefont {Liao}}, \bibinfo {author}
  {\bibfnamefont {J.~L.}\ \bibnamefont {Ellis}}, \bibinfo {author}
  {\bibfnamefont {D.}~\bibnamefont {Zusin}}, \bibinfo {author} {\bibfnamefont
  {C.}~\bibnamefont {Gentry}}, \bibinfo {author} {\bibfnamefont {Q.~L.}\
  \bibnamefont {Nguyen}}, \bibinfo {author} {\bibfnamefont {J.~M.}\
  \bibnamefont {Shaw}}, \bibinfo {author} {\bibfnamefont {A.}~\bibnamefont
  {Pic{\'o}n}}, \bibinfo {author} {\bibfnamefont {L.}~\bibnamefont {Plaja}},
  \bibinfo {author} {\bibfnamefont {H.~C.}\ \bibnamefont {Kapteyn}}, \bibinfo
  {author} {\bibfnamefont {M.~M.}\ \bibnamefont {Murnane}},\ and\ \bibinfo
  {author} {\bibfnamefont {C.}~\bibnamefont {Hern{\'a}ndez-Garc{\'i}a}},\
  }\href {https://doi.org/10.1038/s41566-018-0304-3} {\bibfield  {journal}
  {\bibinfo  {journal} {Nat. Photonics}\ }\textbf {\bibinfo {volume} {13}},\
  \bibinfo {pages} {123} (\bibinfo {year} {2019})}\BibitemShut {NoStop}%
\bibitem [{\citenamefont {Wang}\ \emph {et~al.}(2019)\citenamefont {Wang},
  \citenamefont {Zepf},\ and\ \citenamefont {Rykovanov}}]{Wang2019}%
  \BibitemOpen
  \bibfield  {author} {\bibinfo {author} {\bibfnamefont {J.~W.}\ \bibnamefont
  {Wang}}, \bibinfo {author} {\bibfnamefont {M.}~\bibnamefont {Zepf}},\ and\
  \bibinfo {author} {\bibfnamefont {S.~G.}\ \bibnamefont {Rykovanov}},\ }\href
  {https://doi.org/10.1038/s41467-019-13357-1} {\bibfield  {journal} {\bibinfo
  {journal} {Nat. Commun.}\ }\textbf {\bibinfo {volume} {10}},\ \bibinfo
  {pages} {5554} (\bibinfo {year} {2019})}\BibitemShut {NoStop}%
\bibitem [{\citenamefont {Rebernik Ribi\ifmmode~\check{c}\else \v{c}\fi{}}\
  \emph {et~al.}(2017)\citenamefont {Rebernik Ribi\ifmmode~\check{c}\else
  \v{c}\fi{}}, \citenamefont {R\"osner}, \citenamefont {Gauthier},
  \citenamefont {Allaria}, \citenamefont {D\"oring}, \citenamefont {Foglia},
  \citenamefont {Giannessi}, \citenamefont {Mahne}, \citenamefont {Manfredda},
  \citenamefont {Masciovecchio}, \citenamefont {Mincigrucci}, \citenamefont
  {Mirian}, \citenamefont {Principi}, \citenamefont {Roussel}, \citenamefont
  {Simoncig}, \citenamefont {Spampinati}, \citenamefont {David},\ and\
  \citenamefont {De~Ninno}}]{Rebernik2017}%
  \BibitemOpen
  \bibfield  {author} {\bibinfo {author} {\bibfnamefont {P.~c.~v.}\
  \bibnamefont {Rebernik Ribi\ifmmode~\check{c}\else \v{c}\fi{}}}, \bibinfo
  {author} {\bibfnamefont {B.}~\bibnamefont {R\"osner}}, \bibinfo {author}
  {\bibfnamefont {D.}~\bibnamefont {Gauthier}}, \bibinfo {author}
  {\bibfnamefont {E.}~\bibnamefont {Allaria}}, \bibinfo {author} {\bibfnamefont
  {F.}~\bibnamefont {D\"oring}}, \bibinfo {author} {\bibfnamefont
  {L.}~\bibnamefont {Foglia}}, \bibinfo {author} {\bibfnamefont
  {L.}~\bibnamefont {Giannessi}}, \bibinfo {author} {\bibfnamefont
  {N.}~\bibnamefont {Mahne}}, \bibinfo {author} {\bibfnamefont
  {M.}~\bibnamefont {Manfredda}}, \bibinfo {author} {\bibfnamefont
  {C.}~\bibnamefont {Masciovecchio}}, \bibinfo {author} {\bibfnamefont
  {R.}~\bibnamefont {Mincigrucci}}, \bibinfo {author} {\bibfnamefont
  {N.}~\bibnamefont {Mirian}}, \bibinfo {author} {\bibfnamefont
  {E.}~\bibnamefont {Principi}}, \bibinfo {author} {\bibfnamefont
  {E.}~\bibnamefont {Roussel}}, \bibinfo {author} {\bibfnamefont
  {A.}~\bibnamefont {Simoncig}}, \bibinfo {author} {\bibfnamefont
  {S.}~\bibnamefont {Spampinati}}, \bibinfo {author} {\bibfnamefont
  {C.}~\bibnamefont {David}},\ and\ \bibinfo {author} {\bibfnamefont
  {G.}~\bibnamefont {De~Ninno}},\ }\href
  {https://doi.org/10.1103/PhysRevX.7.031036} {\bibfield  {journal} {\bibinfo
  {journal} {Phys. Rev. X}\ }\textbf {\bibinfo {volume} {7}},\ \bibinfo {pages}
  {031036} (\bibinfo {year} {2017})}\BibitemShut {NoStop}%
\bibitem [{\citenamefont {Vrakking}(2014)}]{Vrakking2014}%
  \BibitemOpen
  \bibfield  {author} {\bibinfo {author} {\bibfnamefont {M.~J.~J.}\
  \bibnamefont {Vrakking}},\ }\href {https://doi.org/10.1039/C3CP53659A}
  {\bibfield  {journal} {\bibinfo  {journal} {Phys. Chem. Chem. Phys.}\
  }\textbf {\bibinfo {volume} {16}},\ \bibinfo {pages} {2775} (\bibinfo {year}
  {2014})}\BibitemShut {NoStop}%
\bibitem [{\citenamefont {Cattaneo}\ \emph {et~al.}(2016)\citenamefont
  {Cattaneo}, \citenamefont {Vos}, \citenamefont {Lucchini}, \citenamefont
  {Gallmann}, \citenamefont {Cirelli},\ and\ \citenamefont
  {Keller}}]{Cattaneo2016}%
  \BibitemOpen
  \bibfield  {author} {\bibinfo {author} {\bibfnamefont {L.}~\bibnamefont
  {Cattaneo}}, \bibinfo {author} {\bibfnamefont {J.}~\bibnamefont {Vos}},
  \bibinfo {author} {\bibfnamefont {M.}~\bibnamefont {Lucchini}}, \bibinfo
  {author} {\bibfnamefont {L.}~\bibnamefont {Gallmann}}, \bibinfo {author}
  {\bibfnamefont {C.}~\bibnamefont {Cirelli}},\ and\ \bibinfo {author}
  {\bibfnamefont {U.}~\bibnamefont {Keller}},\ }\href
  {https://doi.org/10.1364/OE.24.029060} {\bibfield  {journal} {\bibinfo
  {journal} {Opt. Express}\ }\textbf {\bibinfo {volume} {24}},\ \bibinfo
  {pages} {29060} (\bibinfo {year} {2016})}\BibitemShut {NoStop}%
\bibitem [{rep(2020)}]{repo}%
  \BibitemOpen
  \href@noop {} {}\bibinfo {howpublished} {The RMT repository
  {\url{https://gitlab.com/Uk-amor/RMT/rmt}}} (\bibinfo {year}
  {2020})\BibitemShut {NoStop}%
\end{thebibliography}%

\end{document}